\documentclass[11pt,a4paper]{article}
\pdfoutput=1

\usepackage{jheppub}

\usepackage{amssymb}
\usepackage{amsfonts}
\usepackage{graphicx}
\usepackage{epstopdf}
\usepackage{dcolumn}
\usepackage{amsmath}
\usepackage{latexsym,bm}
\usepackage{amsthm}
\usepackage{slashed}
\usepackage{float}
\usepackage{color}
\usepackage{url}
\usepackage{longtable}
\usepackage{subfig}
\usepackage{tikz}
\usepackage{multirow}

\newcommand \xoverline[2][0.75]{
    \sbox{\myboxA}{$\m@th#2$}
    \setbox\myboxB\null
    \ht\myboxB=\ht\myboxA
    \dp\myboxB=\dp\myboxA
    \wd\myboxB=#1\wd\myboxA
    \sbox\myboxB{$\m@th\overline{\copy\myboxB}$}
    \setlength\mylenA{\the\wd\myboxA}
    \addtolength\mylenA{-\the\wd\myboxB}
    \ifdim\wd\myboxB<\wd\myboxA
       \rlap{\hskip 0.5\mylenA\usebox\myboxB}{\usebox\myboxA}%
    \else
        \hskip -0.5\mylenA\rlap{\usebox\myboxA}{\hskip 0.5\mylenA\usebox\myboxB}%
    \fi}
\makeatother

\newcommand{\ba}{\begin{aligned}}
\newcommand{\ea}{\end{aligned}}
\def \be {\begin{equation}}
\def \ee {\end{equation}}
\def \bsp {\begin{split}}
\def \esp {\end{split}}
\def \bea {\begin{eqnarray}}
\def \eea {\end{eqnarray}}

\def\mc{\mathcal}

\def\mb{\mathbb}

\def \bp{\begin{pmatrix}}
\def\ep{\end{pmatrix}}

\def\t{\tilde}

\title{Elliptic Calabi-Yau fivefolds and 2d (0,2) F-theory landscape}

\preprint{\today \hspace*{0.1in} }

\abstract{In this paper, we initiate the study of the 2d F-theory landscape based on compact elliptic Calabi-Yau fivefolds. In particular, we determine the boundary models of the landscape using Calabi-Yau fivefolds with the largest known Hodge numbers $h^{1,1}$ and $h^{4,1}$. The former gives rise to the largest geometric gauge group in the currently known 2d (0,2) supergravity landscape, which is   $E_8^{482\,632\,421}\times F_4^{3\,224\,195\,728}\times G_2^{11\,927\,989\,964}\times SU(2)^{25\,625\,222\,180}$. Besides that, we systematically study the hypersurfaces in weighted projective spaces with small degrees, and check the gravitational anomaly cancellation. Moreover, we also initiate the study of singular bases in 2d F-theory. We find that orbifold singularities on the base fourfold have non-zero contributions to the gravitational anomaly.}

\author[a]{Jiahua Tian}
\author[b]{Yi-Nan Wang}

\affiliation[a]{Abdus Salam International Centre for Theoretical Physics, \\
Strada Costiera 11, 34151, Trieste, Italy}
\affiliation[b]{Mathematical Institute, University of Oxford, \\
Andrew-Wiles Building,  Woodstock Road, Oxford, OX2 6GG, UK}

\emailAdd{jtian@ictp.it}
\emailAdd{yinan.wang@maths.ox.ac.uk}


\begin{document}

\maketitle

\section{Introduction}

In the pursuit of the gobal set of consistent quantum gravity theories, it is very important to identify the boundaries of the string theory landscape, in order to compare them with the swampland bounds~\cite{Vafa:2005ui}. For example, one can ask the following question:

\vspace{0.2cm}

\textit{In a given space-time dimension and amount of supersymmetry, what is the maximal number of fields of a given type in a string compactification model?}

\vspace{0.2cm}

For non-chiral theories with 16 supercharges in $d>3$ space-time dimensions, the maximal rank of gauge group is given by $r_G=26-d$, and it was matched with the swampland bounds~\cite{Kim:2019ths}. 

For theories with eight supercharges, such as 6d $(1,0)$ supergravity, the currently known maximal number of tensor multiplet,  $T=193$, and the maximal rank of the gauge group, $r_G=296$, are both given by F-theory on the elliptic Calabi-Yau threefold $X_3$ with maximal $h^{1,1}$~\cite{Candelas:1997eh,Aspinwall:1997ye,Morrison:2012js,Taylor:2012dr}\footnote{By ``maximal'' we meant the extremal Hodge numbers of Calabi-Yau manifolds as a hypersurface of weighted projective spaces, which appeared in the sequence (3.3) of \cite{Klemm:1996ts}. These numbers represent the records among all the known compact (elliptic) Calabi-Yau manifolds, which are also conjectured to be the rigorous bound in full generality, see~\cite{Taylor:2012dr} for the CY3 case. We will also use this notion of ``maximal'' later on.}:
\be
(h^{1,1},h^{2,1})=(491,11)\,.
\ee
For 5d $\mc{N}=1$ supergravity, the maximal number of vector multiplets is also realized on the same geometry, from the M-theory starting point. These bounds have not been proven as a swampland condition, despite of the presence of worldsheet CFT techniques in these cases~\cite{Heckman:2019bzm,Kim:2019vuc,Lee:2019skh,Katz:2020ewz}.

For theories with four supercharges, such as 4d $\mc{N}=1$ supergravity, the maximal rank of gauge group $r_G=121\,328$ is given by F-theory on the elliptic Calabi-Yau fourfold $X_4$ with maximal known $h^{1,1}$~\cite{Candelas:1997eh,Wang:2020gmi}:
\be
(h^{1,1},h^{2,1},h^{3,1})=(303\,148,0,252)\,.
\ee

The same model also leads to the largest number of axions
\be
N(\rm axion)=181\,820\,.
\ee

On the other hand, F-theory on the mirror Calabi-Yau fourfold with the largest $h^{3,1}$ would lead to the largest number of complex structure moduli and number of flux vacua on a single geometry~\cite{Taylor:2015xtz}.

As a general pattern, the F-theory landscape seems to always provide the answer to the above question in even space-time dimensions. In particular, the point of interest is always the elliptic Calabi-Yau manifold with the largest Hodge numbers.

In this paper, we will extend this logic to the case of 2d (0,2) supergravity with two supercharges, which comes from F-theory on a compact elliptic Calabi-Yau fivefold~\cite{Schafer-Nameki:2016cfr,Lawrie:2016rqe}. As another motivation, the study of (0,2) gauge theories in two dimensions is a rich subject by itself, see e. g.~\cite{Witten:1993yc,Benini:2013xpa,Franco:2015tna,Franco:2016nwv}, and it is interesting to investigate the coupling of a supergravity sector.

In particular, we will study the details of the elliptic Calabi-Yau fivefolds with maximal $h^{1,1}$ or $h^{4,1}$. For the case of maximal $h^{1,1}$:
\be
(h^{1,1},h^{2,1},h^{3,1},h^{4,1},h^{2,2})=(247\,538\,602\,581,0,0,151\,701,758\,522)\,,
\ee
and the 2d (0,2) theory has a geometric gauge group
\be
G=E_8^{482\,632\,421}\times F_4^{3\,224\,195\,728}\times G_2^{11\,927\,989\,964}\times SU(2)^{25\,625\,222\,180}\,.
\ee

The total rank of gauge group is
\be
r_G=66\,239\,044\,388\,,
\ee
which is conjectured to be the largest in the whole 2d (0,2) landscape. 

The construction of the corresponding fourfold base with $h^{1,1}(B_4)=181\,299\,558\,192$ is similar to the 4d case~\cite{Wang:2020gmi}. We tune $E_8$ gauge groups on the toric divisors of a starting point toric fourfold, and then blow up all the non-minimal loci in codimension-two, three and four.

Besides this particular geometric model, we also present the first attempt of studying the set of elliptic Calabi-Yau fivefolds and the 2d F-theory geometric landscape. The constructions of Calabi-Yau fivefolds were explored in~\cite{Kreuzer:2001fu,ahlgren2002points,Haupt:2008nu}, but the elliptic fibration structures have not been discussed in the literature. Namely, we study the Calabi-Yau hypersurfaces of reflexive weighted projective spaces up to degree $d\leq 150$ that have an elliptic fibration structure. For example, the generic fibration over a  ``generalized Hirzebruch fourfold'' is given by a Calabi-Yau hypersurface inside $\mb{P}^{1,1,1,1,n,2n+8,3n+12}$. We also find Calabi-Yau fivefolds with non-zero Hodge numbers $h^{2,1}$ and $h^{3,1}$. The ones with non-zero $h^{3,1}$ describes 2d (0,2) supergravity coupled to 2d Fermi multiplets. The full table of these geometries is listed in Appendix~\ref{app:CY5-list}.

Finally, we checked the 2d gravitational anomaly cancellation conditions~\cite{Lawrie:2016rqe,Weigand:2017gwb} in several cases with or without non-Abelian gauge groups. More interestingly, we also analyzed cases with a singular base, and we found that these orbifold singularities also have a non-zero contribution to the gravitational anomaly.

The structure of this paper is as follows: in section~\ref{sec:2d-F-theory}, we briefly recap the formulation of 2d F-theory and the gravitational anomaly computation. In section~\ref{sec:bound_2d_landscape}, we present the detailed construction of the elliptic Calabi-Yau fivefolds with either largest $h^{1,1}$ or $h^{4,1}$. In section~\ref{sec:CY5s}, we study the geometric structure of a number of other elliptic Calabi-Yau fivefolds. In section~\ref{sec:grav_anomaly}, we check gravitational anomaly cancellation, including the models with a singular base.

\section{Mathematics and physics of 2d F-theory compactifications}
\label{sec:2d-F-theory}

In this section, we introduce the basics of globally consistent compactification of F-theory to $1+1$ dimensions on compact elliptic Calabi-Yau fivefolds, including the geometric tools and the gravitational anomaly computation of the low energy effective theory. In section \ref{sec:basics} we introduce compactification of F-theory on elliptic Calabi-Yau fivefolds with an emphasis on the computation of the massless spectrum of the low energy effective theory. In section \ref{sec:grav_anomaly_intro} we discuss the derivation of gravitational anomaly of the 2d effective theory. The materials in section \ref{sec:basics} and \ref{sec:grav_anomaly_intro} are not new and are all covered in \cite{Schafer-Nameki:2016cfr, Lawrie:2016rqe, Weigand:2017gwb}. In section \ref{sec:construct_cy5} we review the basic toric geometry tools that we will make use of to construct examples of elliptic Calabi-Yau fivefolds.

\subsection{Basic setup of 2d F-theory}\label{sec:basics}

We consider compactification of F-theory on an elliptic Calabi-Yau fivefold $X_5$ whose low energy effective theory is a 2d $\mathcal{N} = (0,2)$ supersymmetric field theory coupled to gravity. In general, an elliptic Calabi-Yau $(n+1)$-fold has the following form:
\begin{equation}
	\begin{array}{ccc}
		\pi: \mathbb{E}_\tau & \rightarrow & Y_{n+1} \\
		 & & \downarrow \\
		 & & B_n
	\end{array}
\end{equation}
and we will mainly focus on the $n = 4$ cases. We further assume that the fibration has a zero section therefore it can be described  by a Weierstrass model:
\begin{equation}
	y^2 = x^3 + fxz^4 + gz^6\,,
\end{equation}
where $f\in\mathcal{O}(-4K_B)$ and $g\in\mathcal{O}(-6K_B)$. Here $K_B$ is the canonical bundle of the base fourfold $B_4$. We will mainly working in the local chart where we can set $z = 1$. Singularities of the elliptic fibration at different codimensions of the base $B_4$ correspond to different physical contents and we list such correspondences in Table \ref{tab1:dim-phys}.
\begin{table}[h]
\begin{center}
	\begin{tabular}{c|c}
		Codimension & Physical data \\
		\hline
		1 & Gauge groups \\
		\hline
		\multirow{2}{*}{2} & Matters in $\mathbf{R}\oplus\overline{\mathbf{R}}$ \\
		 & Bulk-surface matter couplings \\
		\hline
		3 & \multirow{2}{*}{Holomorphic matter couplings} \\
		4 & \\
	\end{tabular}\caption{Singularities and the corresponding physical data of the low energy 2d $\mathcal{N} = (0,2)$ field theory.}\label{tab1:dim-phys}
\end{center}
\end{table}

For our purpose it is sufficient to discuss the codimension-1 and 2 singularities on $B_4$ as we will focus only on the gauge groups and matters in this paper. Codimension-1 singularities are characterized by the vanishing of the discriminant locus:
\begin{align*}
	\Delta = 4f^3 + 27g^2.
\end{align*}
In the IIB physics, the locus $\Delta = 0$ is wrapped by 7-branes, and the gauge group $G_S$ along the codimension-1 locus $S$ is determined by the order of vanishing of $(f, g, \Delta)$ along $S$. The matters are localized at codimension-2 locus of $B_4$ where the order of vanishing of $(f,g,\Delta)$ along $S$ enhances. The matter representations can be determined following Katz-Vafa \cite{Katz:1996xe}. There is also bulk matter that is not localized as we will discuss later. For us it is important to know that with gauge invariant $G_4$ flux, the bulk matter transforms in the adjoint representation of the gauge group $G_S$ and it will contribute to the anomaly.

Besides the 7-branes wrapping codimension-1 loci of $B_4$, there will also be D3-branes wrapping codimension-2 loci of $B_4$ due to tadpole cancellation. The interplay between D3-brane sector and 7-brane sector will also contribute to gravitational anomaly in 2d.

Another indispensable ingredient in the F-theory compactification is the $G_4$ flux which must satisfy the following condition:
\begin{align*}
	G_4 + \frac{1}{2}c_2(X_5)\in H^4(X_5, \mathbb{Z})\cap H^{2,2}(X_5)\,,
\end{align*}
in order for the M-theory compactification on $Y_5$ to preserve two supercharges \cite{Haupt:2008nu}. We will see that $G_4$ flux contributes to the gravitational anomaly from the 3-7 sector.

We will summarize some properties of the the supermultiplets in the 2d $\mathcal{N} = (0,2)$ field theory. They include vector multiplets with one negative chirality complex fermion, chiral multiplets with one positive chirality Weyl fermion, Fermi multiplets with one negative chirality complex fermion and a single gravity multiplet with one positive chirality complex dilatino and one negative chirality gravitino. In 2d there are also tensor multiplets containing real axionic scalar fields arising from KK reduction of the F-theory 4-form field $C_4$. The tensor multiplets will play an important role in the Green-Schwarz mechanism of anomaly cancellation as will be discussed in the next section.

\subsection{Gravitational anomaly cancellation}\label{sec:grav_anomaly_intro}

In 2d the gravitational and gauge anomaly can be described by a gauge invariant polynomial of degree 2 in gauge field strength $F$ and the curvature 2-form $R$:
\begin{equation}
	I_4 = \sum_{\mathbf{R},s}n_s(\mathbf{R})I_s(\mathbf{R})\,,
\end{equation}
where $I_s(\mathbf{R})$ is the anomaly polynomial of a single spin $s$ matter field in representation $\mathbf{R}$ and $n_s(\mathbf{R})$ is the multiplicity of that matter field. 

In general $I_4$ does not have to vanish in a consistent quantum field theory. A gauge variant Green-Schwarz counter-term at tree level can cancel $I_4$ if $I_4$ factorizes suitably. This is possible in 2d because of the existence of an axionic scalar field $c^\alpha$ that gives rise to a self-dual one-form $H^\alpha = dc^\alpha + \Theta^\alpha_iA^i$, such that:
\begin{align*}
	g_{\alpha\beta}*H^\beta = \Omega_{\alpha\beta}H^\beta\,.
\end{align*}
The gauge variant pseudo-action that contains $c^\alpha$ and $H^\alpha$ is:
\begin{equation}
	S_{\text{GS}} = -\frac{1}{4}\int g_{\alpha\beta}H^\alpha\wedge *H^\beta - \frac{1}{2}\int\Omega_{\alpha\beta}c^\alpha\wedge X^\beta\,,
\end{equation}
where $dH^\alpha = X^\alpha$ and $X^\alpha = \Theta^\alpha_iF^i$, $F^i$ is the field strength of the abelian gauge group factor $U(1)_i$. The axionic symmetry of $c^\alpha$ is gauged by $A^i$ with the following transformation rule:
\begin{align*}
	A^i &\rightarrow A^i + d\lambda^i\,, \\
	c^\alpha &\rightarrow c^\alpha - \Theta^\alpha_i\lambda^i\,.
\end{align*}
It is then easy to obtain the gauge variation of $S_{\text{GS}}$ is:
\begin{equation}
	\delta S_{\text{GS}} = \frac{1}{2}\int\Omega_{\alpha\beta}\Theta^\alpha_i\lambda^iX^\beta := 2\pi\int I_{2,\text{GS}}^{(1)}(\lambda)\,.
\end{equation}
Using the descent equations:
\begin{align*}
	I_{4,\text{GS}} = dI_{3,\text{GS}},\ \delta_\lambda I_{4,\text{GS}} = dI_{2,\text{GS}}^{(1)}(\lambda)\,,
\end{align*}
we have:
\begin{equation}
	I_{4,\text{GS}} = \frac{1}{4\pi}\Omega_{\alpha\beta}X^\alpha X^\beta = \frac{1}{4\pi}\Omega_{\alpha\beta}\Theta^\alpha_i\Theta^\beta_j F^i F^j\,.
\end{equation}
We require:
\begin{equation}
	I_4 + I_{4,\text{GS}} = 0\,.
\end{equation}
It is easy to see that since $I_{4,\text{GS}}$ contains only the field strengths of abelian gauge groups, the cancellation is possible only if the gravitational and non-abelian gauge anomalies vanish by themselves and the abelian gauge anomalies factorize suitably. In this paper, we will denote by $I_4$ the gravitational anomaly of the low energy effective theory from a 2d F-theory construction, and we will check if $I_4 = 0$ for a series of examples.

For simplicity we first consider the gravitational sector of F-theory compactification on a smooth Calabi-Yau fivefold $X_5$. Using the duality between F-theory and IIB orientifold we have the following spectrum in the moduli and gravitational sector \cite{Lawrie:2016rqe} in table~\ref{tab:moduli-grav_sector}.
\begin{table}[h]
	\begin{center}
		\begin{tabular}{c|c}
			2d multiplet & Multiplicity \\
			\hline
			Chiral & $h^{2,1}(X_5)+h^{4,1}(X_5)-(-h^{1,1}(B_4)+h^{2,1}(B_4)-h^{3,1}(B_4))-1$ \\
			\hline
			Fermi & $h^{2,1}(B_4)-h^{3,1}(B_4)+h^{3,1}(X_5)$ \\
			\hline
			Tensor & $\tau(B_4)$ \\
			\hline
			Gravity & 1
		\end{tabular}\caption{The 2d supermultiplets in the moduli and gravitational sector of F-theory compactification on $X_5$.}\label{tab:moduli-grav_sector}
	\end{center}
\end{table} 

Here the signature $\tau(B_4)$ is given by
\be
\tau(B_4)=48+2h^{1,1}(B_4)+2h^{3,1}(B_4)-2h^{2,1}(B_4)\,.
\ee

Summing up the contributions of chiral, Fermi and tensor multiplets ($+1$ for chiral multiplets and $(-1)$ for Fermi and tensor multiplets) to the 2d anomaly polynomial we have:
\begin{equation}\label{eq:I_moduli-smooth}
\ba
	I_{4,\text{moduli}}& = \frac{1}{24}p_1(T)(-\tau(B_4) + \chi_1(X_5) - 2\chi_1(B_4))\\
&\equiv\frac{1}{24}p_1(T)\mathcal{A}_{\text{grav}|\text{mod}}\,.
\ea
\end{equation}
where we have used the relation $h^{1,1}(X_5) = 1+h^{1,1}(B_4)$ and the definition of arithmetic genus:
\begin{equation}
	\chi_q(V) = \sum_{p=1}^{\text{dim}V}(-1)^ph^{p,q}(V).
\end{equation}
The gravitational anomaly from the gravity multiplet is:
\begin{equation}
\ba
	I_{4,\text{grav}} &= \frac{1}{24}p_1(T)\times 24\\
&\equiv\frac{1}{24}p_1(T)\mathcal{A}_{\text{grav}|\text{uni}}.
\ea
\end{equation}

We then consider the spectrum of 3-7 sector when a D3 brane wraps genus $g$ curve $C$ in $B_4$. The spectrum is summarized in the table~\ref{tab:37sector}.
\begin{table}[h]
	\begin{center}
		\begin{tabular}{c|c}
			Multiplet & Multiplicity \\
			\hline
			Chiral & $h^0(C,N_{C/B_4}) + g - 1 + c_1(B_4)\cdot C$ \\
			\hline
			Fermi & $h^0(C,N_{C/B_4}) + g - 1 + 7c_1(B_4)\cdot C$
		\end{tabular}\caption{The 2d supermultiplets in the 3-7 sector.}\label{tab:37sector}
	\end{center}
\end{table}

Summing up the contributions from chiral and Fermi multiplets (note again they have opposite contributions), we have:
\begin{equation}
\ba
	I_{4,3-7} &= \frac{1}{24}p_1(T)(-6c_1(B_4)\cdot C)\\
&\equiv\frac{1}{24}p_1(T)\mathcal{A}_{\text{grav}|\text{3-7}}\,.
\ea
\end{equation}

The various arithmetic genus above can be computed via index theorem and we have:
\begin{align}
	\chi_1(B_4) &= \frac{1}{180}\int_{B_4}(-31c_4-11c_1c_3+3c_2^2+4c_1^2c_2-c_1^4)\,, \\
	\chi_1(X_5) &= \int_{B_4}(90c_1^4 + 3c_1^2c_2 - \frac{1}{2}c_1c_3)\,, \\
	\tau(B_4) &= \frac{1}{180}\int_{B_4}(12c_2^2 - 56c_1c_3 + 56c_4 - 4c_1^4 + 16c_1^2c_2)\,.
\end{align}
Here $c_i$ is the $i^{\text{th}}$ Chern class of the base $B_4$. For a smooth Calabi-Yau fivefold we have:
\begin{equation}
	[C] = \frac{1}{24}\pi_*c_4(X_5) = 15c_1^3 + \frac{1}{2}c_1c_2\,.
\end{equation}
Here $\pi:X_5\rightarrow B_4$ is the fibration map, and $\pi_*$ is the push forward map from $X_5$ to $B_4$. 

Summing up all the contributions we have:
\begin{equation}
	I_4 = I_{4,\text{moduli}} + I_{4,\text{grav}} + I_{4,3-7} = \frac{1}{24}p_1(T)(-24\chi_0(B_4) + 24)
\end{equation}
where:
\begin{equation}
	\chi_0(B_4) = \frac{1}{720}\int_{B_4}(-c_4 + c_1c_3 + 3c_2^2 + 4c_1^2c_2 - c_1^4)\,.
\end{equation}
Recall that for a base $B_4$ to support a smooth elliptic fibration for a Calabi-Yau fivefold, we have $h^{0,0}(B_4) = 1$ and $h_{k,0}(B_4) = 0$ for $k\neq0$. Therefore $\chi_0(B_4) = 1$ and the gravitational anomaly is cancelled for smooth elliptic Calabi-Yau fivefolds.

We now assume that the fibration contains non-abelian gauge groups from 7-branes and charged 7-7 matters. In addition we turn on $G_4$ flux. In this case the terms above needs slight modification ad there will be a new term $I_{4,7-7}$ contributing to the gravitational anomaly from the 7-brane sector.

Suppose that the divisor $S\subset B_4$ is wrapped by 7-branes. The Kodaira fiber is singular over $S$ and the Calabi-Yau fivefold $X_5$ is singular. We assume that the singular $X_5$ admits a crepant resolution $\tilde{f}:\tilde{X}_5\rightarrow X_5$ and $G_4\in H^{2,2}_{\text{vert}}(\tilde{X}_5)$. In this situation the $\chi(X_5)$ term in $I_{4,\text{moduli}}$ (\ref{eq:I_moduli-smooth}) is replaced by $\chi(\tilde{X}_5)$. The D3-brane class $[C]$ is corrected to:
\begin{equation}
	[C] = \frac{1}{24}\pi_*c_4(\tilde{X}_5) - \frac{1}{2}\pi_*(G_4\cdot G_4)\,.
\end{equation}
The anomaly polynomial from the non-trivial 7-brane sector is:
\begin{equation}
\ba
	I_{4,7-7} &=\frac{1}{24}p_1(T)\left[ \sum_{\mathbf{R}}\text{dim}(\mathbf{R})\chi(\mathbf{R}) - \text{rk}(G)\chi(\mathbf{adj})\right]\\
&\cong\frac{1}{24}p_1(T)\mathcal{A}_{\text{grav}|\text{7-7}}
\,.
\ea
\end{equation}
In section \ref{sec:grav_anomaly}, we investigate cases with only non-Higgsable gauge groups and $\chi(\mathbf{adj})$ is purely geometric. To cancel the gravitational anomaly the following relation must hold:
\begin{equation}\label{eq:total_grav_anom}
	\mathcal{A}_{\text{grav}|\text{mod}}+\mathcal{A}_{\text{grav}|\text{uni}}+\mathcal{A}_{\text{grav}|\text{3-7}}+\mathcal{A}_{\text{grav}|\text{7-7}} = 0\,.
\end{equation}
The above equation puts a set of topological constraints that every crepant resolution $\tilde{X}_5\rightarrow X_5$ with consistent background $G_4$ flux on $\tilde{X}_5$ must satisfy. It will be verified on a set of Calabi-Yau fivefolds $\tilde{X}_5$ in section \ref{sec:grav_anomaly}.

\subsection{Construction of Calabi-Yau fivefold hypersurfaces}\label{sec:construct_cy5}

In this section, we will review some basics tools of toric geometry that we will use to construct Calabi-Yau fivefolds as hypersurfaces in toric sixfolds. The techniques are standard and can be found in \cite{cox2011toric}. We will use Batyrev's construction \cite{batyrev1993dual} to construct Calabi-Yau hypersurfaces in a reflexive polytope. We will explain the details in a moment.

We will start with an $(n+1)$-d reflexive polytope $\Delta$ in an $(n+1)$-d lattice in $M_{\mathbb{R}}$. That is, $\Delta\subset M_{\mathbb{R}}$ contains $\mathbf{0}$ and both $\Delta$ and $\Delta^*$ are lattice polytopes where $\Delta^*\subset N_{\mathbb{R}}$ is defined as:
\begin{align*}
	\Delta^*:=\{ v\in N_{\mathbb{R}}:\ \langle u,v\rangle \geq -1, \forall u\in\Delta \}\,,
\end{align*}
where $N_{\mathbb{R}}$ is the dual lattice of $M_{\mathbb{R}}$.

The polytope $\Delta^*$ defines a toric fan $\Sigma$ and to each point $v_i$ on the boundary of $\Delta^*$ one can associate a homogeneous coordinate $z_i$. We denote by $Y_{n+1}$ the $(n+1)$-d toric variety defined by $\Sigma$. To each point $u_i\in\Delta$ one can associate a monomial $m_i = \prod_j z_j^{\langle u_i,v_j \rangle + 1}$. The locus $\sum_i a_im_i = 0$ ($a_i$ are generic non-vanishing complex coefficients) defines a hypersurface $X_n\subset Y_{n+1}$ in the anticanonical class $-K_{Y_{n+1}}$ of $Y_{n+1}$. Therefore $X_{n}$ is a Calabi-Yau $n$-fold. Note that there is no guarantee that $X_{n}$ is smooth when $n>3$.

For the Calabi-Yau $n$-fold hypersurface $X_n$ defined from the reflexive pair $(\Delta^*,\Delta)$, the (stringy) Hodge numbers can be computed with the Batyrev formula \cite{batyrev1993dual,batyrev1997stringy}:
\be
h^{1,1}(X_n)=l(\Delta^*)-(n+2)-\sum_{\mathrm{dim}\Theta^*=n}l'(\Theta^*)+\sum_{\mathrm{dim}\Theta^*=n-1}l'(\Theta^*)l'(\Theta)\label{Batyrevh11}
\ee
\be
h^{m,1}(X_n)=\sum_{\mathrm{dim}\Theta^*=n-m}l'(\Theta^*)l'(\Theta)\quad (1< m< n-1)\label{Batyrevhm1}
\ee
\be
h^{n-1,1}(X_n)=l(\Delta)-(n+2)-\sum_{\mathrm{dim}\Theta=n}l'(\Theta)+\sum_{\mathrm{dim}\Theta=n-1}l'(\Theta)l'(\Theta^*)\label{Batyrevhd1}
\ee
Here $\Theta^*$ and $\Theta$ means the faces on $\Delta^*$ and $\Delta$ respectively. $l(.)$ means the number of integral points in a polytope, and $l'(.)$ means the number of interior points on a face. 

For the cases we will discuss in this paper, they are all $n$-d hypersurfaces defined in some $(n+1)$-d ambient toric varieties that are also elliptically fibered over some $(n-1)$-d bases. Such a fibration structure can be easily read off by studying the toric fans of their ambient toric varieties. For all the examples in this paper, after a suitable $SL(6,\mathbb{Z})$ transformation, the vertices of $\Delta^*$ can be put into the following form:
\begin{align*}
	&\tilde{v}_1 = (0,0,0,0,0,1),\ \tilde{v}_2 = (0,0,0,0,1,0),\ \tilde{v}_3 = (0,0,0,1,-2,-3), \\
	&\tilde{v}_{i+3}=(v_i,-2,-3)\,.
\end{align*}
This is of the form introduced in \cite{Candelas:1996su} and is known to be a $\mathbb{P}^{2,3,1}$ fibration over a base toric variety $B_4$. The fan of $B_4$ has toric rays $v_i$, and we denote the convex hull of it  by the polytope $\Delta_{B_4}$.

The Calabi-Yau hypersurface defined by the pair ($\Delta^*,\Delta$) is thus an elliptic fibration over $B_4$. Note that to fully specify the toric variety corresponding to $\Delta^*$, a triangulation is also required. We require the triangulation to be fine (uses all the points in $\Delta^*$), regular (resulting variety is projective and K\"ahler) and star (the simplices define the cones of a toric fan). Though a triangulation of $\Delta_{B_4}$ is needed to compute some detailed geometrical data such as intersection numbers on $B_4$, the computation of the Hodge numbers and the characteristic classes of $B_4$ depends only on the rays in the fan $\Sigma_{B_4}$ associated with $\Delta_{B_4}$. Therefore in later sections where we compute Hodge numbers and characteristic classes of $B_4$ and $Y_5$, we will choose a convenient triangulation to facilitate our computations and the results are indeed independent from our choices. 

The base varieties of the examples in Section \ref{sec:grav_anomaly} are particularly easy in this sense since their triangulations are unique. In contrast, the triangulations of the bases of the examples in Section \ref{sec:bound_2d_landscape} are far from being unique, but one does not need to worry about any specific choice of triangulation since we will be computing Hodge numbers only and the key data involed this computation are the numbers of cones in various codimensions which are constants across all fine-star-regular triangulations (FRST). 

For example, if the elliptic fibration does not have codimension-two ord$(f,g)\geq (4,6)$, codimension-three ord$(f,g)\geq (8,12)$ or codimension-four ord$(f,g)\geq (12,18)$ non-minimal loci, then we expect the Shioda-Tate-Wazir formula to hold, independent of the triangulation of the base:
\be
h^{1,1}(X_5)=h^{1,1}(B_4)+\mathrm{rk}(G)+1\,,\label{STW}
\ee
where $G$ is the 2d geometric gauge group.

For most examples in our paper with $E_8$ geometric gauge groups, we will try to construct a smooth base $B_4$ that supports a flat fibration. To do that, we will first pick all the primitive rays $\rho$ inside $\Delta_{B_4}$ and this we will denote by $S$ this set of primitive rays. We will denote by $B_{\text{toric}}$ the toric variety given by $S$ (and a suitable triangulation of it). We then pick the subset $S_{E_8}\subset S$ whose elements are the rays that correspond to divisor supporting Kodaira $II^*$ fiber, that is, carrying an $E_8$ gauge group. To find these rays we consider the following two polytopes:
\begin{align*}
	&\Delta_F = \{u\in\mathbb{Z}^4|\langle u,v_i\rangle + 4 \geq 0\ ,\ \forall v_i\in S \}, \\
	&\Delta_G = \{u\in\mathbb{Z}^4|\langle u,v_i\rangle + 6 \geq 0\ ,\ \forall v_i\in S \}.
\end{align*}
The points in $\Delta_F$ correspond to monomials in the class $-4K_{B_{\text{toric}}}$ and the points in $\Delta_G$ correspond to monomials in the class $-6K_{B_{\text{toric}}}$. The orders of vanishing of the polynomials $f\in O(-4K_{B_{\text{toric}}})$ and $g\in O(-6K_{B_{\text{toric}}})$ in the Weierstrass model along a divisor $D_i$ corresponding to the primitive ray $u_i\in S$ are:
\begin{align*}
	&\text{ord}_{D_i}(f) = \text{min}_{u\in \Delta_F} (\langle u,v_i\rangle + 4), \\
	&\text{ord}_{D_i}(g) = \text{min}_{u\in \Delta_G} (\langle u,v_i\rangle + 6), \\
\end{align*}
We denote by $S_{E_8}$ the set of $v_i$'s such that $\text{ord}_{D_i}(f) = 4$ and $\text{ord}_{D_i}(g) = 5$.

Usually the set $S_{E_8}$ does not give rise to a compact base and we need to add several rays manually. After adding these rays by hand we arrive at a base we call $B_{\text{seed}}$. This base needs to be blown-up to be free from codimension-two $(4,6)$ locus, codimension-three $(8,12)$ and codimension-four $(12,18)$ non-minimal loci. Focusing on $S_{E_8}$, we can compute the number of 4d cones in $S_{E_8}$, $n_{4D}$. By assigning a convenient triangulation to $S_{E_8}$ we can then compute the number of 3d and 2d cones in $S_{E_8}$, $n_{3D}$ and $n_{2D}$ respectively and $n_{1D}$ is simply the number of rays in $S_{E_8}$. Note that the $n_{4D}$, $n_{3D}$, $n_{2D}$ and $n_{1D}$ are all indeed independent of triangulation and our choice is simply to make the computation easier. There is the following correspondence between those numbers and the gauge web structure over $S_{E_8}$:
\begin{table}[h]
	\begin{center}
		\begin{tabular}{c|c}
			 & Number of \\
			\hline
			$n_{4D}$ & $(E_8,E_8,E_8,E_8)$ point \\
			\hline
			$n_{3D}$ & $(E_8,E_8,E_8)$ curve \\
			\hline
			$n_{2D}$ & $(E_8,E_8)$ surface \\
			\hline
			$n_{1D}$ & $E_8$ divisor
		\end{tabular}
	\end{center}
\end{table}

For each of the above intersecting $E_8$ structure there is a sequence of blow-ups one needs to perform over $B_{\text{seed}}$ to finally arrive at a smooth base $B_4$. We will present the process in the Appendix~\ref{sec:4-E8-blp}.

\section{The boundaries of 2d (0,2) F-theory landscape}\label{sec:bound_2d_landscape}

\subsection{Calabi-Yau $d$-fold with extremal Hodge numbers}

We first compute the ambient reflexive polytope for Calabi-Yau $d$-fold with extremal Hodge numbers, which is a generalization of the sequence (3.3) in~\cite{Klemm:1996ts}. We first define a sequence of integers $m_k$, with
\be
m_0=1\ ,\ m_{k+1}=m_k(m_k+1)\,.
\ee
The first a few $m_i$ are
\be
m_1=2\ ,\ m_2=6\ ,\ m_3=42\ ,\ m_4=1\,806\ ,\ m_5=3\,263\,442\,.
\ee
Then the ambient reflexive polytope is a $(d+1)$-dimensional weighted projective space $\mb{P}^{1,1,d_1,d_2,\dots,d_{d}}$. The weights are computed as:
\be
d_1=2\cdot m_{d-1}\ ,\ d_2=(2+d_1)\cdot m_{d-2}\ ,\ d_{k+1}=\left(2+\sum_{i=1}^k d_i\right)\cdot m_{d-k-1}\,.
\ee

For the elliptic CY3 $X_3$ with $(h^{1,1},h^{2,1})=(11,491)$, the ambient weighted projective space is $\mb{P}^{1,1,12,28,42}$.

For the elliptic CY4 $X_4$ with $(h^{1,1},h^{2,1},h^{3,1})=(252,0,303\,148)$, the ambient weighted projective space is $\mb{P}^{1,1,84,516,1204,1806}$.

For the elliptic CY5 $X_5$ with the largest $h^{4,1}$, from the rules above, we expect the ambient weighted projective space to be $\mb{P}^{1,1,3612,151\,788,932\,412,2\,175\,628,3\,263\,442}$.

Using the terminologies in section \ref{sec:construct_cy5}, a weighted projective space $\mb{P}^{1,w_1,\dots,w_{d+1}}$ corresponds to an ambient polytope $\Delta^*$ with vertices: 
\be
\begin{array}{c}
v_1=(0,\dots,0,1)\\
v_2=(0,\dots,1,0)\\
 \vdots\\
v_{d+1}=(1,0,\dots,0)\\
v_{d+2}=(-w_1,-w_2,\dots,-w_{d+1})
\end{array}
\ee

As one can check, the pairs $(\Delta^*,\Delta)$ above are all reflexive.

\subsection{Maximal $h^{4,1}$}
\label{sec:max-h41}

In this section, we construct the Calabi-Yau fivefold $X_5$ with the largest $h^{4,1}$ from the reflexive pair $(\Delta^*,\Delta)$, where $\Delta^*$ corresponds to $\mb{P}^{1,1,3\,612,151\,788,932\,412,2\,175\,628,3\,263\,442}$. We will explicitly construct the elliptic fibration structure and the base fourfold $B_4$.

The weighted projective space $\mb{P}^{1,1,3\,612,151\,788,932\,412,2\,175\,628,3\,263\,442}$ has the following vertices:
\be
\ba
&\t v_1=(0,0,0,0,0,1)\ ,\ \t v_2=(0,0,0,0,1,0)\ ,\ \t v_3=(0,0,0,1,0,0),\cr
&\t v_4=(0,0,1,0,0,0)\ ,\ \t v_5=(0,1,0,0,0,0)\ ,\ \t v_6=(1,0,0,0,0,0),\cr
&\t v_7=(-1,-3\,612,-151\,788,-932\,412,-2\,175\,628,-3\,263\,442)\,.\label{v-CY5}
\ea
\ee
Its dual polytope has the following vertices:
\be
\ba
&\t u_1=(-1,-1,-1,-1,-1,-1)\ ,\ \t u_2=(-1,-1,-1,-1,-1,1),\cr
&\t u_3=(-1,-1,-1,-1,2,-1)\ ,\ \t u_4=(-1,-1,-1,6,-1,-1),\cr
&\t u_5=(-1,-1,42,-1,-1,-1)\ ,\ \t u_6=(-1,1\,806,-1,-1,-1,-1),\cr
&\t u_7=(6\,526\,883,-1,-1,-1,-1,-1)\,.\label{u-CY5}
\ea
\ee

From the Batyrev formula, one can compute $h^{1,1}(X_5)=151701$. The last term in (\ref{Batyrevh11}) vanishes. Similarly, $h^{2,1}(X_5)$ and $h^{3,1}(X_5)$ both vanishes as well.

The other Hodge numbers can be computed by Landau-Ginzburg methods~\cite{Vafa:1989xc}:
\be
h^{4,1}(X_5)=247\,538\,602\,581\ ,\ h^{2,3}(X_5)=2\,722\,923\,718\,202\ ,\ h^{2,2}(X_5)=758\,522\,.
\ee
They satisfy the relation~\cite{Haupt:2008nu}:
\be
11h^{1,1}-10h^{2,1}-h^{2,2}+h^{2,3}+10h^{3,1}-11h^{4,1}=0
\ee

We perform an $SL(6,\mb{Z})$ rotation on $v_i$:
\be
\label{SL6rotation}
M=\begin{pmatrix} 
1 & 0 & 0 & 0 & 0 & 0\\
0 & 1 & 0 & 0 & 0 & 0\\
0 & 0 & 1 & 0 & 0 & 0\\
0 & 0 & 0 & 1 & 0 & 0\\
-2 & -2 & -2 & -2 & 1 & 0\\
-3 & -3 & -3 & -3 & 0 & 1
\end{pmatrix}
\ee

The resulting vertices are
\be
\ba
&\t v_1'=(0,0,0,0,0,1)\ ,\ \t v_2'=(0,0,0,0,1,0)\ ,\ \t v_3'=(0,0,0,1,-2,-3),\cr
&\t v_4'=(0,0,1,0,-2,-3)\ ,\ \t v_5'=(0,1,0,0,-2 -3)\ ,\ \t v_6'=(1,0,0,0,-2,-3),\cr
&\t v_7'=(-1,-3\,612,-151\,788,-932\,412,-2,-3)\,.
\ea
\ee
Hence it is in form of $\mb{P}^{1,2,3}$ bundle over a 4d base $B_4$, whose 4d polytope $\Delta_{B_4}$ has the following vertices:
\be
\Delta_{B_4}=\{(1,0,0,0), (0,1,0,0), (0,0,1,0), (0,0,0,1), (-1,-3\,612,-151\,788,-932\,412)\}.\label{smallh11-B4-pol}
\ee
The base $B_4$ of $X_5$ is a $B_3$ fibration over $\mb{P}^1$. $B_3$ is exactly the threefold base for the elliptic CY4 $X_4$ with $h^{1,1}=h^{3,1}=151\,700$, as similar phenomenon is observed in the lower dimensional case~\cite{Taylor:2015xtz}. Note that $X_4$ has an elliptic fibration with geometric gauge groups~\cite{Candelas:1997eh}
\be
G_{\rm 4d}=E_8^{1\,285}\times F_4^{3\,792}\times G_2^{10\,092}\times SU(2)^{15\,108}\,.
\ee

 To construct the rays and cones on $B_4$ and $B_3$. We first compute the set of lattice points $\{x,y,z,w\}$ in the polytope (\ref{smallh11-B4-pol}), with the following condition:
\be
\mathrm{gcd}(x,y,z,w)=1\,.
\ee

Among these points, we select the ones that correspond to divisors with $E_8$ gauge group, which form the set $S_{E_8}$. Such a point $v$ satisfy the following condition:
\be
\mathrm{min}_{u\in\Delta_G}(\langle u,v\rangle+6)=5\,,
\ee
where the $\Delta_G$ polytope is the set of lattice points $u=(u_x,u_y,u_z,u_w)$ satisfying
\be
u_x\geq -6\ ,\ u_y\geq -6\ ,\ u_z\geq -6\ ,\ u_w\geq -6\ ,\ -u_x-3612u_y-151\,788u_z-932\,412u_w\geq -6\,.
\ee
It turns out that there are 1\,285 points satisfying the conditions, and they are all in the form of $(0,y,z,w)$. Then we can construct a non-compact toric threefold $B_{E_8}^{(3)}$ with the 3d rays $(y,z,w)$. After a triangulation, we find that there are 2\,508 $(E_8,E_8,E_8)$ 3d cones and 3\,792 $(E_8,E_8)$ 2d cones on $B_{E_8}^{(3)}$. Then we add three additional rays $(1,0,0)$, $(0,1,0)$, $(0,0,1)$ and a number of additional 3d cones into $B_{E_8}^{(3)}$, such that the resulting base is a compact one $B_{\rm seed}^{(3)}$. 


Finally, after we blow up the $(E_8,E_8,E_8)$ 3d cones and $(E_8,E_8)$ 2d cones according to~\cite{Wang:2020gmi} (also see  appendix \ref{sec:4-E8-blp}), we get a base $B_{\rm toric}^{(3)}$ with 90\,652 rays and $h^{1,1}(B_{\rm toric}^{(3)})=90\,649$. The number of rays is computed as follows. We start with the $B_{\rm seed}^{(3)}$ with 1\,288 rays. Then for each of the 2\,508 $(E_8,E_8,E_8)$ 3d cones, we need to add 19 additional rays in the interior. For each of the 3\,792 $(E_8,E_8)$ 2d cones, we need to add 11 additional rays on it. Thus these numbers add up to 90\,652. Finally, to get the base $B_3$, we checked that there are 310 $E_8$ divisors $p$ on $B_{\rm toric}$ with non-toric $(4,6)$-curves. This can be checked by the following criterion:
\be
|\{u\in\Delta_G|\langle u,p\rangle+6=5\}|>1\,.\label{g5-crit}
\ee
It turns out that all of these non-toric $(4,6)$-curves are irreducible. After these curves are blown up, we get the non-toric base $B_3$ with $h^{1,1}(B_3)=90\,959$.

After adding up the rank of geometric gauge group, we get exactly the following Shioda-Tate-Wazir formula in CY4 case:
\be
h^{1,1}(X_4)=h^{1,1}(B_3)+\mathrm{rk}(G)+1=151\,700\,.
\ee

Then the 4d base $B_4$ is constructed as $B_3$ fibered over $\mb{P}^1$ with the addition of two rays $(1,0,0,0)$ and $(-1,-3\,612,-151\,788,-932\,412)$. The geometric gauge group on $B_4$ remains the same, and there is no additional base locus to be blow up. Thus the base $B_4$ has $h^{1,1}(B_4)=90\,960$, and we have exactly
\be
h^{1,1}(X_5)=h^{1,1}(B_4)+\mathrm{rk}(G)+1=151\,701\,.
\ee

For the 2d F-theory on $X_5$, the geometric gauge group is also
\be
G=E_8^{1\,285}\times F_4^{3\,792}\times G_2^{10\,092}\times SU(2)^{15\,108}\,.
\ee

\subsection{Maximal $h^{1,1}$}

In this section, we construct Calabi-Yau fivefold $X_5$ with the largest $h^{1,1}$, along with its elliptic fibration structure.

We take (\ref{u-CY5}), and perform an $SL(6,\mb{Z})$ rotation:
\be
\label{SL6rotation-2}
M=\begin{pmatrix} 
1 & 0 & 0 & 0 & 2 & 3\\
0 & 1 & 0 & 0 & 2 & 3\\
0 & 0 & 1 & 0 & 2 & 3\\
0 & 0 & 0 & 1 & 2 & 3\\
0 & 0 & 0 & 0 & 1 & 1\\
0 & 0 & 0 & 0 & 1 & 2
\end{pmatrix}
\ee

The resulting vertices are
\be
\ba
&\t v_1'=(-6,-6,-6,-6,-2,-3)\ ,\ \t v_2'=(0,0,0,0,0,1)\ ,\ \t v_3'=(0,0,0,0,1,0),\cr
&\t v_4'=(-6,-6,-6,1,-2,-3)\ ,\ \t v_5'=(-6,-6,37,-6,-2, -3),\cr
&\t v_6'=(-6,1\,801,-6,-6,-2,-3)\ ,\ \t v_7'=(6\,526\,878,-6,-6,-6,-2,-3)\,.
\ea
\ee
Naively, it is in form of $\mb{P}^{1,2,3}$ bundle over a 4d base $B_4$ with vertices
\be
\ba
\Delta_{B_4}=&\{(-6,-6,-6,-6), (-6,-6,-6,1), (-6,-6,37,-6), (-6,1\,801,-6,-6),\cr
& (6\,526\,878,-6,-6,-6)\}.
\ea
\ee

Nonetheless, the vertices such as $(-6,-6,-6,-6)$ cannot correspond to a ray on a smooth base $B_4$, because all the coordinates are dividable by four. It should be interpreted as six times the ray $(-1,-1,-1,-1)$ on $B_4$, which carries an $E_8$ gauge group.

\begin{figure}
\begin{center}
\includegraphics[height=7cm]{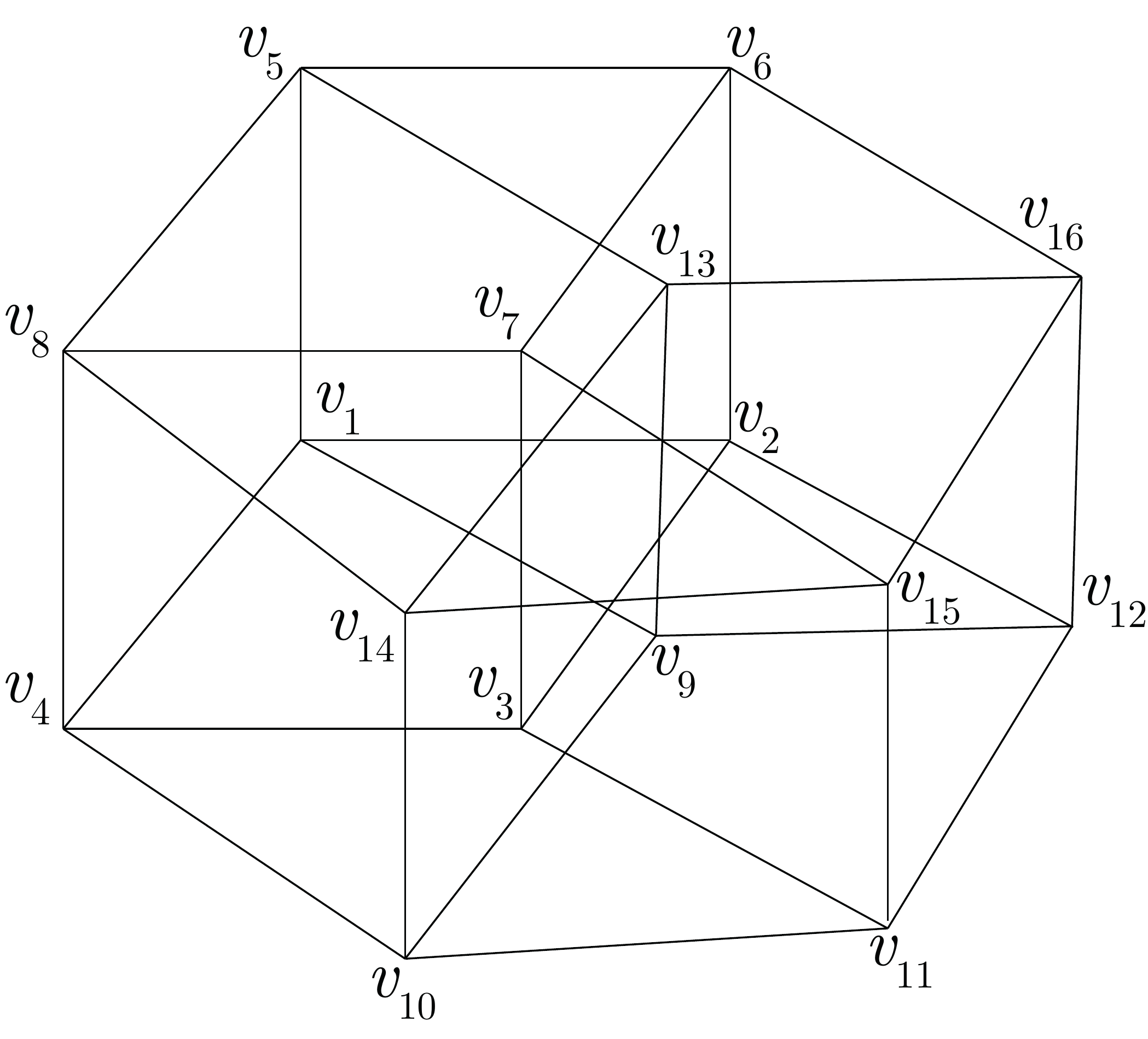}
\caption{The vertices of hyper truncated pyramid $\Delta_{E_8}$, for the elliptic Calabi-Yau fivefold with the largest $h^{1,1}$.}\label{f:hyperpyramid}
\end{center}
\end{figure}

Now we write down the set of rays $S_{E_8}$ whose corresponding toric divisor supports $E_8$ gauge algebra:
\begin{align}
	S_{E_8} = &\{(x,y,z,-1)| -1\leq z\leq 6,\ -1\leq y\leq \frac{932\,412-151\,788z}{3\,612},\nonumber\\
     &-1\leq x\leq 932\,413-151\,788z-3\,612y\}\ \nonumber \\
	&\cup\ \{(x,y,-1,0)|-1\leq y\leq 42,\ -1\leq x\leq 151\,789-3\,612y\}\ \nonumber \\
	&\cup\ \{(x,-1,0,0)|-1\leq x\leq 3\,613\}\ \nonumber \\
	&\cup\ \{(1,0,0,0),(-1,0,0,0)\}.\label{eq:SE8}
\end{align}
There are in total
\be
n_{1D}=482\,632\,421
\ee
integral points in this set. Now we are going to construct the non-compact toric fourfold $B_{E_8}$ with rays in the set $S_{E_8}$. We denote by $\Delta_{E_8}$ the convex hull polytope of $S_{E_8}$. $\Delta_{E_8}$ has a shape of hyper truncated pyramid, with the following 16 vertices, see figure~\ref{f:hyperpyramid}:
\begin{align}
	& v_1 = (-1,-1,-1,-1),\ v_2 = (-1,300,-1,-1),\ v_3 = (601,300,-1,-1),\nonumber \\
	& v_4 = (1\,087\,813,-1,-1,-1),\ v_5 = (-1,-1,6,-1),\ v_6 = (-1,6,6,-1),\nonumber \\
	& v_7 = (13,6,6,-1),\ v_8 = (25\,297,-1,6,-1),\label{eq:d1vert} \\
	& v_9 = (-1,-1,-1,0),\ v_{10} = (155\,401,-1,-1,0),\ v_{11} = (85,42,-1,0),\nonumber\\
     & v_{12} = (-1,42,-1,0),\label{eq:d2vert} \\
	& v_{13} = (-1,-1,0,0),\ v_{14} = (3\,613,-1,0,0),\label{eq:d3vert} \\
	& v_{15} = (1,0,0,0),\ v_{16} = (-1,0,0,0)
\end{align}

We observe that the vertices of $\Delta_{E_8}$ can be naturally organized in the following manner: the vertices in (\ref{eq:d1vert}) are the 8 vertices of the first line of (\ref{eq:SE8}), the vertices in ( \ref{eq:d2vert}) are the 4 vertices of the second line of (\ref{eq:SE8}) and the vertices in (\ref{eq:d3vert}) are the 2 ends of the third line of (\ref{eq:SE8}). 

It is a fact that the number of simplicial 4d cones is independent of the choice of triangulation of the 4d fan given by the primitive rays in $S_{E_8}$. To compute the number of simplicial 4d cones, we only need to calculate the volume of the $\Delta_{E_8}$, which turns out to be
\begin{align*}
	\text{vol}(\Delta_{E_8}) = 114\,084\,800.
\end{align*}
Therefore the number of simplicial 4d cones is:
\begin{align}
\label{maxh11-n4D}
	n_{4D} = 4!\times\text{vol}(\Delta_{E_8}) = 2\,738\,035\,200.
\end{align}

To compute the total number of 3d cones on $B_{E_8}$, one can use the following trick. On a compact toric fourfold, each 4d cone contains four 3d cones, while each 3d cone is shared by two 4d cones. Hence the number of 3d cones on a compact toric fourfold should be the twice of the number of 4d cones. However, the base $B_{E_8}$ is non-compact, with the following boundary 2d faces:
\be
\ba
\label{maxh11-boundary2d}
&v_2 v_3 v_6 v_7, v_5 v_6 v_7 v_8, v_9 v_{10} v_{11} v_{12}, v_{10} v_{11}v_{14} v_{15}, v_9 v_{10}v_{13}v_{14}, v_9 v_{12}v_{13}v_{16}, v_2 v_3 v_{11}v_{12},\cr
&v_2 v_6 v_{12} v_{16}, v_3 v_7 v_{11} v_{15}, v_5 v_6 v_{13} v_{16}, v_5 v_8 v_{13} v_{14}, v_7 v_8 v_{14} v _{15}\,.
\ea
\ee
In the above list, we take the 2d faces inside a single 3d face with non-zero contribution to the 4d volume of $\Delta_{E_8}$.

Now one takes two times the number of 4d cones (\ref{maxh11-n4D}), plus additional 3d cones from the boundary set (\ref{maxh11-boundary2d}) divided by two. We get
\be
\ba
\label{maxh11-n3D}
n_{3D}&=2n_{4D}+\frac{1}{2}\times 7\,056\,216\cr
&=5\,479\,598\,508\,.
\ea
\ee

Then to compute the number of 2d cones on $B_{E_8}$, one needs to carefully add up all the contributions from each faces of $\Delta_{E_8}$. The result is
\be
\label{maxh11-n2D}
n_{2D}=3\,224\,195\,728\,.
\ee

With the number of 4d, 3d and 2d cones, we now construct the base $B_4$ by blowing up the $(E_8,E_8,E_8,E_8)$, $(E_8,E_8,E_8)$ and $(E_8,E_8)$ collisions, according to section~\ref{sec:4-E8-blp}. For each 4d cone, there are in total 15 exceptional divisor in the interior after blowing up the $(E_8,E_8,E_8,E_8)$ collision. For each 3d cone and 2d cone, there are in total 19 and 11 exceptional divisors, respectively. Finally, there are a number of non-toric blow ups on the divisors on $B_{E_8}$. They can be checked by the criterion  (\ref{g5-crit}) in this case as well, and there are in total
\be
N_{\rm non-toric}=167\,873\,112
\ee
of these divisors (which are all irreducible). In this whole process, we are only blowing up loci where $(4,6)\leq \mathrm{ord}(f,g)<(8,12)$ at codimension-two, $(8,12)\leq \mathrm{ord}(f,g)<(12,18)$ at codimension-three and $(12,18)\leq\mathrm{ord}(f,g)<(16,24)$ at codimension-four. Hence the number of complex structure moduli of $X_5$ is unchanged and it is still within a finite distance of the moduli space.

Finally, we need to add the rays $(-6,-6,-6,1)$, $(-6,-6,37,-6)$ and $(-6,1\,801,-6,-6)$ back into the base, to make $B_4$ compact. The total $h^{1,1}(B_4)$ is then
\be
\ba
h^{1,1}(B_4)&=n_{1D}+15n_{4D}+19n_{3D}+11n_{2D}+N_{\rm non-toric}+3-4\cr
&=181\,299\,558\,192\,.
\ea
\ee

To compute the $h^{1,1}(X_5)$ of this elliptic Calabi-Yau fivefold. We add the rank of non-Higgsable gauge groups: for each 4d cone, there is a single $SU(2)$; for each 3d cone, the additional gauge group is $G_2\times SU(2)^3$; for each 2d cone, the additional gauge group is $F_4\times G_2^2\times SU(2)^2$; for each $E_8$ ray, the gauge rank is 8.

Thus we have (\ref{STW})
\be
\ba
h^{1,1}(X_5)&=h^{1,1}(B_4)+8n_{1D}+n_{4D}+5n_{3D}+10n_{2D}+1\cr
&=247\,538\,602\,581\,.
\ea
\ee

This number is exactly the same as the $h^{4,1}$ of its mirror in section~\ref{sec:max-h41}. Hence the elliptic fibration structure is completely correct.

The numbers of each type of gauge groups are
\be
\ba
n(E_8)&=n_{1D}\cr
&=482\,632\,421\,,\cr
n(F_4)&=n_{2D}\cr
&=322\,419\,5728\,,\cr
n(G_2)&=n_{3D}+2n_{2D}\cr
&=11\,927\,989\,964\,,\cr
n(SU(2))&=n_{4D}+3n_{3D}+2n_{2D}\cr
&=25\,625\,222\,180\,.
\ea
\ee

The total 2d geometric gauge group is
\be
G=E_8^{482\,632\,421}\times F_4^{3\,224\,195\,728}\times G_2^{11\,927\,989\,964}\times SU(2)^{25\,625\,222\,180}\,.
\ee

\section{Various elliptic Calabi-Yau fivefolds}
\label{sec:CY5s}

In this section, we explicitly study a number of elliptic Calabi-Yau fivefolds as hypersurfaces of $\mb{P}^{1,w_1,w_2,w_3,w_4,w_5,w_6}$, which constructed from a reflexive polytope. While the full list for $\sum_{i=1}^6 w_i<150$ is presented in Appendix~\ref{app:CY5-list}, we will discuss a few examples in full detail and explain the origin of the non-vanishing Hodge numbers $h^{2,1}(X_5)$ and $h^{3,1}(X_5)$.

\subsection{Hypersurface of $\mb{P}^{1,1,1,1,n,2n+8,3n+12}$}\label{sec:CYseries}

In these section, we consider ambient spaces in form of $\mb{P}^{1,1,1,1,n,2n+8,3n+12}$, $n\in\mb{Z}_+$. For $n\geq 4$, the toric base fourfold is a ``generalized Hirzebruch fourfold'' $B_{n,4}$. In general, it is a toric fourfold with $h^{1,1}(B_{n,4})=2$ and it has the structure of a $\mb{P}^1$ fibration over $\mb{P}^3$. The fan of $B_{n,4}$ has the following rays
\begin{align*}
	&v_1= (1,0,0,0),\ v_2 = (0,1,0,0),\ v_3 = (0,0,1,0),\ v_4 = (0,0,0,1), \\
	&v_5 = (-1,-1,-1,-n),\ v_6 = (0,0,0,-1)\,.
\end{align*}
The list of 4d cones of the toric variety is complete:
\begin{align*}
	&(1,2,3,4),\ (1,2,3,6),\ (1,2,4,5),\ (1,2,5,6), \\
	&(1,3,4,5),\ (1,3,5,6),\ (2,3,4,5),\ (2,3,5,6)\,,
\end{align*}
where $(i,j,k,l)$ denotes the 4D cone whose rays are $v_i$, $v_j$, $v_k$ and $v_l$. These bases have
\be
\chi(B_{n,4})=8\,.
\ee

For $1\leq n\leq 3$, the base fourfold is a weighted projective space $\mb{P}^{1,1,1,1,n}$. The rays are
\begin{align*}
	&v_1 = (1,0,0,0),\ v_2 = (0,1,0,0),\ v_3 = (0,0,1,0),\ v_4 = (0,0,0,1), \\
	&v_5 = (-1,-1,-1,-n)\,.
\end{align*}
The list of 4d cones is
\be
(1,2,3,4),\ (1,2,3,5),\ (1,2,4,5),\ (1,3,4,5),\ (2,3,4,5)\,.
\ee

For $\mathbb{P}^{1,1,1,1,n,2n+8.3n+12}$ to be reflexive, $n$ can only take the following values:
\begin{align}
	n = 1, 2, 3, 4, 6, 8, 12, 24.
\end{align}
The data of $X_5$ for these cases are summarized in the table~\ref{t:Bn}.
\begin{table}
	\begin{center}
		\begin{tabular}{|c|c|c|}
			\hline
			$n$ & Gauge group & $(h^{1,1},h^{2,1},h^{3,1},h^{4,1},h^{2,3})$ \\
			\hline
			1 & None & (2, 0, 0, 56\,977, 626\,727) \\
			\hline
			2 & None & (2, 0, 0, 59\,054, 649\,574) \\
			\hline
			3 & None & (2, 0, 0, 72\,888, 801\,751) \\
			\hline
			4 & None & (3, 1, 0, 93\,190, 1\,025\,070) \\
			\hline
			6 & $SU(3)$ & (5, 0, 0, 151\,471, 1\,666\,132) \\
			\hline
			8 & $SO(8)$ & (7, 0, 0, 235\,299, 2\,588\,220) \\
			\hline
			12 & $E_6$ & (9, 0, 0, 494\,933, 5\,444\,174) \\
			\hline
			24 & $E_8$ & (11, 0, 0, 2\,314\,879, 25\,463\,560) \\
			\hline
		\end{tabular}
            \caption{The Hodge numbers of the generic elliptic CY5 over a smooth base $B_n$. The non-Higgsable gauge group is also listed.}\label{t:Bn}
	\end{center}
\end{table}

For $n=1$, the Calabi-Yau fivefold $X_5$ is a generic ellptic fibration over $\mb{P}^4$. The fibration is smooth, and the Hodge numbers are
\be
(h^{1,1},h^{2,1},h^{3,1},h^{4,1})=(2,0,0,56\,977)\,.
\ee

For $n=2$, $X_5$ is a generic fibration over the weighted projective space $\mb{P}^{1,1,1,1,2}$. The base $\mb{P}^{1,1,1,1,2}$ has a codimension-four $\mb{C}^4/\mb{Z}_2$ orbifold singularity at the intersection point $v_1 v_2 v_3 v_5=D_1\cdot D_2\cdot D_3\cdot D_5$. Similarly, the Calabi-Yau fivefold also has a codimension-four terminal singularity over this point. From the Batyrev formula, the Hodge numbers are different from the generic fibration over $\mb{P}^4$:

\be
(h^{1,1},h^{2,1},h^{3,1},h^{4,1})=(2,0,0,59\,054)\,.
\ee

For $n=3$, similarly $X_5$ is a generic fibration over the weighted projective space $\mb{P}^{1,1,1,1,3}$, with a $\mb{C}^4/\mb{Z}_3$ orbifold singularity at the intersection point $v_1 v_2 v_3 v_5=D_1\cdot D_2\cdot D_3\cdot D_5$. The Hodge numbers are:

\be
(h^{1,1},h^{2,1},h^{3,1},h^{4,1})=(2,0,0,72\,888)\,.
\ee

For $n=4$, $X_5$ is a generic fibration over a generalized Hirzebruch fourfold $B_{4,4}$. There is no gauge group on $X_5$, and the Hodge numbers are
\be
(h^{1,1},h^{2,1},h^{3,1},h^{4,1})=(3,1,0,93\,190)\,.
\ee

$h^{1,1}(X_5)$ exactly matches (\ref{STW}), and there is a non-zero $h^{2,1}(X_5)=1$. The harmonic $(2,1)$-form is constructed as follows. The normal bundle and canonical bundle of the divisor $D_6$ corresponding to $v_6=(0,0,0,-1)$ satisfies
\be
N_{D_6}=K_{D_6}\,.
\ee 
Hence the base is locally Calabi-Yau near the divisor $D_6$. Then the elliptic fiber over $D_6$ is a smooth toric $T^2$ with a constant modulus $\tau$. Now we take the $(1,0)$ form of this $T^2$ and wedge it with the Poincar\'{e} dual of $D_6$ (a $(1,1)$-form). Thus we get an a contribution to $h^{2,1}$. This divisor is similar to a single $(-2)$-curve on the base in the cases of elliptic CY3, which also has an additional contribution to $h^{2,1}$ of the CY3~\cite{Morrison:2012js}\footnote{We thank Andreas Braun and Washington Taylor for the discussions here, in an unfinished project before.}.

For $n=6$, $X_5$ is a generic fibration over the generalized Hirzebruch fourfold $B_{6,4}$. There is a type $IV_s$ singular fiber on $D_6$ with an $SU(3)$ gauge group, the Hodge numbers are
\be
(h^{1,1},h^{2,1},h^{3,1},h^{4,1})=(5,0,0,151\,471)\,.
\ee
We can check that $h^{1,1}(X_5)=h^{1,1}(B_{6,4})+\mathrm{rk}(SU(3))+1$.

For $n=8$, $X_5$ is a generic fibration over the generalized Hirzebruch threefold $B_{8,4}$. There is a type $I_{0,s}^*$ singular fiber on $D_6$ with an $SO(8)$ gauge group, the Hodge numbers are
\be
(h^{1,1},h^{2,1},h^{3,1},h^{4,1})=(7,0,0,235\,299)\,.
\ee
Hence we have $h^{1,1}(X_5)=h^{1,1}(B_{8,4})+\mathrm{rk}(SO(8))+1$.

For $n=12$, $X_5$ is a generic fibration over the generalized Hirzebruch threefold $B_{12,4}$. There is a type $IV_{s}^*$ singular fiber on $D_6$ with an $E_6$ gauge group, the Hodge numbers are
\be
(h^{1,1},h^{2,1},h^{3,1},h^{4,1})=(9,0,0,494\,933)\,.
\ee
Hence $h^{1,1}(X_5)=h^{1,1}(B_{12,4})+\mathrm{rk}(E_6)+1$.

For $n=24$, $X_5$ is a generic fibration over the generalized Hirzebruch threefold $B_{24,4}$. There is a type $II^*$ singular fiber on $D_6$ with an $E_8$ gauge group. The Hodge numbers are
\be
(h^{1,1},h^{2,1},h^{3,1},h^{4,1})=(11,0,0,2\,314\,879)\,.
\ee
Hence $h^{1,1}(X_5)=h^{1,1}(B_{24,4})+\mathrm{rk}(E_8)+1$.

In other dimensions, there also exists a similar series of elliptically fibered Calabi-Yau $(d+1)$-dimensional hypersurfaces $X_{d+1}$ in $(d+2)$-dimensional ambient weighted projective spaces. Consider a $(d+2)$-dimensional weighted projective space $W_{n,d} = \mb{P}^{1,1,\cdots,1,n,2(n+d),3(n+d)}$, for its corresponding polyhedron to be reflexive, $n$ can only take the values $6d$ and its divisors. For $n\geq d$,  the Calabi-Yau hypersurface $X_{d+1}$ in $W_{n,d}$ is elliptically fibered over a toric base $\mb{F}_n^{(d)}$ with the following rays:
\begin{align*}
	v_1 &= (1,0,\cdots,0), \\
	v_2 &= (0,1,\cdots,0), \\
	&\phantom{b=\,} \vdots \\
	v_d &= (0,0,\cdots,1), \\
	v_e &= (-1,-1,\cdots,-n), \\
	v_g &= (0,0,\cdots,-1).
\end{align*}
The triangulation of $\Delta_{B_{n,d}}$ is:
\begin{align*}
	&(1,2,\cdots,d),\ (1,2,\cdots,d-1,g),\\
	&(1,2,\cdots,d-2,d,e),\ (1,2,\cdots,d-2,e,g), \\
	&(1,2,\cdots,d-3,d-1,d,e),\ (1,2,\cdots,d-3,d-1,e,g), \\
	&\phantom{b=\,} \vdots \\
	&(1,3,4,\cdots,d-1,d,e),\ (1,3,4,\cdots,d-1,e,g), \\
	&(2,3,\cdots,d-1,d,e),\ (2,3,\cdots,d-1,e,g).
\end{align*}
and we have $\chi(B_{n,d}) = 2d$.

Note that when $d$ is even the largest four divisors of $n_{\text{max}} = 6d$ are $6d$, $3d$, $2d$ and $\frac{3}{2}d$ and when $d$ is odd the largest four divisors of $n_{\text{max}} = 6d$ are $6d$, $3d$, $2d$ and $d$ (or $\frac{6}{5}d$ depends on whether $5|d$). For $n = 6d$, there is $E_8$ gauge group along $D_g$. For $n = 3d$, there is $E_6$ group along $D_g$. For $n = 2d$, there is $SO(8)$ group along $D_g$. When $d$ is even, for $n = \frac{3}{2}d$, there is $SU(3)$ along $D_g$ and for $n < \frac{3}{2}d$ there is no gauge group on $B_{n,d}$. When $d$ is odd, for $n \leq d$ (or $n \leq \frac{6}{5}d$), there is no gauge group on $B_{n,d}$. 

The most well-known case of this series is when $d = 2$. The Hirzebruch surfaces $\mb{F}_3$, $\mb{F}_4$, $\mb{F}_6$ and $\mb{F}_{12}$ carry $SU(3)$, $SO(8)$, $E_6$ and $E_8$ non-Higgsable gauge groups respectively. The series in 3d, known as generalized Hirzebruch threefolds, has also been explored in literatures~\cite{Mohri:1997uk,Taylor:2017yqr}. Note that here $d = 3$ is odd. As $n = 3$ is the fourth largest divisor of $n_{\text{max}} = 18$, there is no gauge group on the base $B_{3,3}$ which is the generalized Hirzebruch threefold $\tilde{\mathbb{F}}_3$.

\subsection{An example with non-zero $h^{2,1}$ and $h^{3,1}$: $(7, 3, 171, 53\,192)$}

Here the Calabi-Yau fivefold $X_5$ is the degree 120 hypersurface in $\mathbb{P}^{1,3,3,3,10,40,60}$. $X_5$ is a $\mathbb{P}^{2,3,1}$ fibration over the base given by the FRST of the polytope $\Delta_{B_4}$ whose rays are listed in table~\ref{t:7-3-171-53192}.
\begin{table}[h]
	\begin{center}
		\begin{tabular}{c|c}
			$v_1$ & (1,0,0,0) \\
			\hline
			$v_2$ & (0,1,0,0) \\
			\hline
			$v_3$ & (0,0,1,0) \\
			\hline
			$v_4$ & (0,0,0,1) \\
			\hline
			$v_5$ & (0,0,0,-1) \\
			\hline
			$v_6$ & (-1,-1,-1,-3) \\
			\hline
			$v_7$ & (-1,-1,-1,-4) \\
			\hline
			$v_8$ & (-2-2,-2,-7) \\
			\hline
			$v_9$ & (-3,-3,-3,-10)
		\end{tabular}
	\end{center}
\caption{The rays on the toric fourfold base $B_4$ of the elliptic Calabi-Yau fourfold in $\mb{P}^{1,3,3,3,10,40,60}$, with Hodge numbers $(h^{1,1},h^{2,1},h^{3,1},h^{4,1})=(7, 3, 171, 53\,192)$.}\label{t:7-3-171-53192}
\end{table}

$\Delta_{B_4}$ is small enough such that a concrete triangulation can be easily found. We triangulate $\Delta_{B_4}$ by giving the 4D cones as follows:
\begin{align*}
	&(1,2,3,4), (1,2,3,5), (1,2,4,6), (1,2,5,7), (1,2,6,9), (1,2,8,9), (1,2,7,8), \\
	&(1,3,4,6), (1,3,5,7), (1,3,6,9), (1,3,8,9), (1,3,7,8), (2,3,4,6), (2,3,5,7), \\
	&(2,3,6,9), (2,3,8,9), (2,3,7,8)\,,
\end{align*}
where $(i,j,k,l)$ denotes the 4d cone whose rays are $v_i$, $v_j$, $v_k$ and $v_l$. There is an $SU(2)$ gauge group on the divisor $D_6$ corresponding to the ray $v_6$. We can compute
\begin{align}
	\chi(B_4) = \int_{B_4}c_4 = 17\,.
\end{align}

In this case, the non-zero $h^{2,1}(X_5)$ can be explained similar to the case of generic fibration on $\mb{F}_4^{(4)}$. The divisor $D_5$ has normal bundle $N_{D_5}=K_{D_5}$, which can be checked from
\be
v_5=\frac{1}{4}(v_1+v_2+v_3+v_7)\,.
\ee
$v_1$, $v_2$, $v_3$ and $v_7$ are neighbors of $v_5$. Hence there is a harmonic $(2,1)$-form, which is constructed from wedging the Poincar\'{e} dual $(1,1)$-form of $D_5$ with the $(1,0)$-form on the constant torus over $D_5$.

Similar thing happens for $D_7$ and $D_8$, as the rays satisfy
\be
\ba
v_7&=\frac{1}{2}(v_5+v_8)\,,\cr
v_8&=\frac{1}{2}(v_7+v_9)\,.
\ea
\ee

In total, there are three harmonic $(2,1)$-form of $X_5$ constructed in this way, which matches $h^{2,1}(X_5)=3$.

The non-zero $h^{3,1}(X_5)$ is explained in another way. Denote the base coordinates of $B_4$ by $z_1,z_2,\dots,z_9$. The local Tate model near the divisor $D_6$ with $SU(2)$ is \cite{Bershadsky:1996nh}
\be
y^2+b_3 z_6y+b_6 z_6^2=x^3+b_1 z_6 xy+b_2 z_6 x^2 +b_4 z_6^2 x \,.
\ee

We have
\be
b_3=F_{20}(z_1,z_2,z_3)\ ,\ b_6=F_{40}(z_1,z_2,z_3)\,.
\ee

Here $F_i$ is generic homogeneous polynomial of degree $i$. Note that the coefficients $b_3$ and $b_6$ of Tate model do not depend on $z_4$ and $z_9$, although $D_4$ and $D_9$ intersect $D_6$. Thus $b_3$ and $b_6$ can be thought as sections of line bundles on $\mb{P}^2\times\mb{P}^1$. The coordinates of $\mb{P}^2$ are $z_1,z_2,z_3$ and the coordinates of $\mb{P}^1$ are $z_4$ and $z_9$.

After the resolution $(x,y,z_6;\delta_1)$ \cite{Lawrie:2012gg}\footnote{It means the replacement $(x,y,z_6)\rightarrow (x\delta_1,y\delta_1,z_6\delta_1)$ followed by the dividing the equation by $\delta_1^2$. The exceptional divisor is given by the equation $\delta_1=0$.}, the equation is transformed into
\be
y^2+F_{20}(x_1,x_2,x_3)z_6y+F_{40}(z_1,z_2,z_3)z_6^2=(x^3+b_1  z_6 xy+b_2 z_6 x^2 +b_4  z_6^2 x)\delta_1\,.
\ee
The exceptional divisor $\delta_1=0$ has equation
\be
y^2+F_{20}(z_1,z_2,z_3)z_6y+F_{40}(z_1,z_2,z_3)z_6^2=0\,.
\ee

Note that if one set $z_6=1$, then the equation
\be
y^2+F_{20}(z_1,z_2,z_3)y+F_{40}(z_1,z_2,z_3)=0
\ee
is a complex surface $S$ with the following Newton polytope:
\be
\Delta_3=\{(0,0,2),(0,0,0),(40,0,0),(0,40,0)\}\,.
\ee
This Newton polytope has 171 interior points, hence $S$ has $h^{2,0}(S)=171$. Taking into account the coordinate $z_6$, $z_4$ and $z_9$, the whole topology of the exceptional divisor $\delta_1=0$ should be $S\times\mb{P}^1\times\mb{P}^1$. Wedging the non-trivial $(2,0)$-form of $S$ with the Poincar\'{e} dual $(1,1)$-form of $\delta_1=0$, we get 171 $(3,1)$-forms in $X_5$, which exactly matches $h^{3,1}(X_5)$.

\subsection{An example with a large $h^{3,1}(X_5)$: $(11,0,2\,024,28\,575)$}

Here we study an elliptic Calabi-Yau fivefold with a large $h^{3,1}(X_5)$. We take the toric ambient space to be the weighted projective space $\mb{P}^{1,6,6,6,6,50,75}$. The Hodge numbers are
\be
(h^{1,1}(X_5),h^{2,1}(X_5),h^{3,1}(X_5),h^{4,1}(X_5))=(11,0,2\,024,28\,575)\,.
\ee
After the $SL(6,\mb{Z})$ rotation (\ref{SL6rotation}), the vertices of the 6d reflexive polytope $\Delta^*$ are
\be
\ba
\t v_1&=(0,0,0,0,0,1)\cr
\t v_2&=(0,0,0,0,1,0)\cr
\t v_3&=(0,0,0,1,-2,-3)\cr
\t v_4&=(0,0,1,0,-2,-3)\cr
\t v_5&=(0,1,0,0,-2,-3)\cr
\t v_6&=(1,0,0,0,-2,-3)\cr
\t v_7&=(-6,-6,-6,-6,-2,-3)\cr
\ea
\ee

The vertices of the 4d base polytope $\Delta_{B_4}$ are:
\be
\ba
v_1&=(1,0,0,0)\cr
v_2&=(0,1,0,0)\cr
v_3&=(0,0,1,0)\cr
v_4&=(0,0,0,1)\cr
v_5&=(-6,-6,-6,-6)
\ea
\ee
The vertex $v_5$ is a multiple of six. Hence one can speculate that the elliptic Calabi-Yau fivefold is an elliptic fibration over $\mb{P}^4$, with type $II^*$ Kodaira fiber on the ray $(-1,-1,-1,-1)$ (tuned $E_8$ gauge group). We label the corresponding divisor of the rays of $\mb{P}^4$ as follows:
\be
\ba
&(1,0,0,0):\ z_1=0\ ,\ (0,1,0,0):\ z_2=0\ ,\ (0,0,1,0):\ z_3=0\ ,\ \cr
&(0,0,0,1):\ z_4=0\ ,\ (-1,-1,-1,-1):\ z_5=0\,.
\ea
\ee

The Calabi-Yau hypersurface equation can be read off from the lattice points in the polytope $\Delta$, which is the dual polytope of $\Delta^*$. The vertices are:
\be
\ba
\label{E8-mirror}
\t u_1&=(-6,-6,-6,-6,-1,-1)\cr
\t u_2&=(19,-6,-6,-6,-1,-1)\cr
\t u_3&=(-6,19,-6,-6,-1,-1)\cr
\t u_4&=(-6,-6,19,-6,-1,-1)\cr
\t u_5&=(-6,-6,-6,19,-1,-1)\cr
\t u_6&=(0,0,0,0,2,-1)\cr
\t u_7&=(0,0,0,0,-1,1)\,.
\ea
\ee 

The Tate model of $X_5$ can be written as:
\be
\ba
&y^2+F_4(z_1,z_2,z_3,z_4,z_5)z_5 xy+F_{12}(z_1,z_2,z_3,z_4,z_5)z_5^3y\cr
=&x^3+F_8(z_1,z_2,z_3,z_4,z_5)z_5^2 x^2+F_{16}(z_1,z_2,z_3,z_4,z_5)z_5^4 x+F_{25}(z_1,z_2,z_3,z_4,z_5)z_5^5\,,
\ea
\ee
and the Weiertrass model can be written as:
\be
y^2=x^3+F_{16}(z_1,z_2,z_3,z_4,z_5)z_5^4 x+F_{25}(z_1,z_2,z_3,z_4,z_5)z_5^5\,.
\ee

Here $F_i(z_1,z_2,z_3,z_4,z_5)$ are generic homogeneous polynomials of degree $i$ in the variables $z_1,z_2,z_3,z_4,z_5$. As one can see, the Weierstrass $g$ polynomial has the following expansion around $u=0$:
\be
g=F_{25}(z_1,z_2,z_3,z_4)z_5^5+\mc{O}(u^6)\,.
\ee
We need to blow up the non-minimal codimension-two $(4,6)$ locus at $z_5=F_{25}(z_1,z_2,z_3,z_4)=0$, which describes a Fermat surface with degree 25.

As a consequence, the new non-toric base fourfold $B_4$ has $h^{3,1}=2024$. The reason is that the Fermat surface $F_{25}(v,w,s,t)=0$ has the following Hodge numbers (see e. g. \cite{schutt2010lines}):
\be
h^{i,j}=\bp 1 & 0 & 2\,024\\
0 & 9\,225 & 0\\
2\,024 & 0 & 1\ep
\ee
Especially, the Hodge number $h^{2,0}=2\,024$. The harmonic $(2,0)$-forms on the Fermat surface, wedged with the $(1,1)$-form which is the Poincar\'{e} dual of the $E_8$ divisor, give rise to $(3,1)$-forms on the base $B_4$. Hence we have
\be
h^{3,1}(B_4)=2\,024\,,
\ee
which can be again uplifted to the $h^{3,1}(X_5)=2\,024$.

On the other hand, the value of $h^{1,1}(X_5)$ matches (\ref{STW}):
\be
h^{1,1}(X_5)=h^{1,1}(B_4)+\mathrm{rank}(G)+1=11\,.
\ee
Here $h^{1,1}(B_4)=2$ after the single blow up along the non-toric fermat surface, and the gauge group rank is 8.

\subsection{An example with a large $h^{2,1}(X_5)$: $(28\,575,2\,024,0,11)$}

To construct the mirror Calabi-Yau fivefold of the $X_5$ in the last section, we take the vertices (\ref{E8-mirror}) and perform the $SL(6,\mb{Z})$ rotation (\ref{SL6rotation-2})

We get the vertices
\be
\ba
\t v_1&=(-6,-6,-6,-6,-2,-3)\cr
\t v_2&=(19,-6,-6,-6,-2,-3)\cr
\t v_3&=(-6,19,-6,-6,-2,-3)\cr
\t v_4&=(-6,-6,19,-6,-2,-3)\cr
\t v_5&=(-6,-6,-6,19,-2,-3)\cr
\t v_6&=(0,0,0,0,1,0)\cr
\t v_7&=(0,0,0,0,0,1)\,.
\label{2024h21-vertices}
\ea
\ee
Therefore $X_5$ is the Calabi-Yau hypersurface in a $\mathbb{P}^{2,3,1}$ bundle fibered over the base variety associated with the 4d polytope $\Delta_{B_4}$ with the vertices:
\begin{align}
	\Delta_{B_4} =& \{(-6,-6,-6,-6),(19,-6,-6,-6),\nonumber \\
	&\ (-6,19,-6,-6),(-6,-6,19,-6),(-6,-6,-6,19)\}.
\end{align}
$\Delta_{B_4}$ is generated by 21437 primitive vectors. The rays in the set $S_{E_8}$ are given by the non-zero vectors in the polytope $\Delta_{E_8}$ whose vertices are:
\begin{align}
	\Delta_{E_8} =& \{(-1,-1,-1,-1),(3,-1,-1,-1),\nonumber \\
	&\ (-1,3,-1,-1),(-1,-1,3,-1),(-1,-1,-1,3)\}. 
\end{align}
We have $|S_{E_8}| = 69$ and $\text{vol}(\Delta_{E_8}) = \frac{32}{3}$. The numbers of $n$-dimensional cones in $\Delta_{S_{E_8}}$ with an arbitrary FRST are:
\begin{align}
	&n_{4D} = 4!\times\text{vol}(\Delta_{E_8}) = 256, \nonumber \\
	&n_{3D} = 2n_{4D} + \frac{1}{2}\times 64 = 544, \nonumber \\
	&n_{2D} = 356.
\end{align}

In this case, the numbers of gauge groups can be explicitly worked out due to the relatively small number of rays in $\Delta_4$. The numbers of rays that support different gauge groups are:
\begin{align}
	n(E_8) = 69,\ n(F_4) = 356,\ n(G_2) = 1\,256,\ n(SU(2)) = 2\,600.
\end{align}
In addition, there are also 5712 primitive rays carry type $II$ fiber. With the extra 53 non-toric blow-ups, we can compute $h^{11}$ to be:
\begin{align}
	h^{11} =&\ 21437 - 4 + 69\times\text{rank}(E_8) + 356\times\text{rank}(F_4)\nonumber \\
	&\ + 1\,256\times\text{rank}(G_2) + 2\,600\times\text{rank}(SU(2)) + 53 + 1\nonumber \\
	=&\ 28\,575.
\end{align}

The numbers of gauge groups can also be used to derive the numbers of $n$-dimensional cones of the polytope $\Delta_{E_8}$. Using the results of blow-ups of $\Delta_{E_8}$ at different codimensions that were worked out in Section \ref{sec:4-E8-blp}, we have:
\begin{align}
	n(E_8) = n_{\text{rays}},\ n(F_4) = n_{2D},\ n(G_2) = 2n_{2D} + n_{3D},\ n(SU(2)) = 2n_{2D} + 3n_{3D} + n_{4D}.
\end{align}
Therefore we obtain $n_{\text{rays}} = 69$, $n_{4D} = 256$, $n_{3D} = 544$ and $n_{2D} = 356$ which match our computation using only the combinatorial data of the polytope $\Delta_{E_8}$.

For the non-trivial $h^{2,1}(X_5)$, it can be explained as following. Consider the 3d face in the base polytope $\Delta_4$ with vertices $(19,-6,-6,-6)$, $(-6,19,-6,-6)$, $(-6,-6,19,-6)$ and $(-6,-6,-6,19)$. This face has 2\,024 interior points, which corresponds to 2024 base divisor with a locally trivial elliptic fibration (constant $\tau$). Then for each base divisor $D$, we can construct a $(2,1)$-form in terms of the Poincar\'{e} dual (1,1)-form of $D$ wedge the $(1,0)$-form on the constant torus. There are in total 2024 of them. 

This is also consistent with the computation of $h^{2,1}(X_5)$ using Batyrev formula. The dual face of $(v_2,v_3,v_4,v_5)$ in (\ref{2024h21-vertices}) is a 2d face with vertices $(-6,-6,-6,-6,-1,-1)$, $(0,0,0,0,-1,1)$, $(0,0,0,0,2,-1)$, which exactly has one interior point $(-1,-1,-1,-1,0,0)$. Thus we can compute $h^{2,1}(X_5)=2024$ from (\ref{Batyrevhm1}).

\section{Gravitational anomaly cancellation in 2d}\label{sec:grav_anomaly}

\subsection{Smooth base with non-Higgsable gauge groups}

As introduced in section~\ref{sec:grav_anomaly_intro}, to cancel the gravitational anomaly, the following expression needs to vanish:
\begin{equation}
	I_{\text{grav}} = \frac{1}{24}p_1(T)(\mathcal{A}_{\text{grav}|\text{7-7}}+\mathcal{A}_{\text{grav}|\text{mod}}+\mathcal{A}_{\text{grav}|\text{uni}}+\mathcal{A}_{\text{grav}|\text{3-7}})
\end{equation}
where
\begin{align}
	\mathcal{A}_{\text{grav}|\text{7-7}} &= \sum_{\mathbf{R}}\text{dim}(\mathbf{R})\chi(\mathbf{R}) - \text{rk}(G)\chi(\mathbf{adj}), \\
	\mathcal{A}_{\text{grav}|\text{mod}} &= -\tau(B_4) + \chi_1(\tilde{X}_5) - 2\chi_1(B_4), \\
	\mathcal{A}_{\text{grav}|\text{uni}} &= 24, \\
	\mathcal{A}_{\text{grav}|\text{3-7}} &= -6c_1(B_4)\cdot\bigg(\frac{1}{24}\pi_*(c_4(\tilde{X}_5)) - \frac{1}{2}\pi_*(G_4\cdot G_4)\bigg) \label{eq:A37}.
\end{align}

For a smooth base, the Hirzebruch signature $\tau(B_4)$ and Chern character $\chi_1(B_4)$ can be computed by \cite{Klemm:1996ts}
\be
\label{tauB4}
\tau(B_4)=\frac{1}{180}\left(12c_2^2-56c_1 c_3+56 c_4-4c_1^4+16 c_1^2 c_2\right)
\ee
\be
\label{chi1B4}
\chi_1(B_4)=\frac{1}{180}\left(-31 c_4-14c_1 c_3+3c_2^2+4c_1^2 c_2-c_1^4\right)\,.
\ee

We compute $\pi_*(c_4(\tilde{X}_5))$ with at most a single non-abelian gauge group supported on a base divisor whose class is $S$ and list them in table~\ref{t:Cs}. Part of them were already computed in \cite{Schafer-Nameki:2016cfr,Weigand:2017gwb}, and these formula can also be found in an analogous computation for the elliptic Calabi-Yau fourfolds in \cite{Esole:2017kyr,Grimm:2009yu}.
\begin{table}[h]
\begin{center}
\begin{tabular}{c|c}
	Gauge group & $\pi_*(c_4(\hat{X}_5))$ \\
	\hline
	- & $C_s$ \\
	\hline
	$SU(2)$ & $C_s - 294c_1^2S + 84c_1S^2 - 6S^3$ \\
	\hline
	$SU(3)$ & $C_s - 456c_1^2S + 192c_1S^2 - 24S^3$ \\
	\hline
	$SU(4)$ & $C_s - 600c_1^2S + 336c_1S^2 - 60S^3$ \\
	\hline
	$SU(5)$ & $C_s - 750c_1^2S + 525c_1S^2 - 160S^3$ \\
	\hline
	$SU(6)$ & $C_s - 894c_1^2S + 753c_1S^2 - 210S^3$ \\
	\hline
	$SO(8)$ & $C_s - 648c_1^2S + 384c_1S^2 - 72S^3$ \\
	\hline
	$SO(10)$ & $C_s - 756c_1^2S + 528c_1S^2 - 120S^3$ \\
	\hline
	$G_2$ & $C_s - 456c_1^2S + 192c_1S^2 - 24S^3$ \\
	\hline
	$F_4$ & $C_s - 648c_1^2S + 384c_1S^2 - 72S^3$ \\
	\hline
	$E_6$ & $C_s - 774c_1^2S + 549c_1S^2 - 126S^3$ \\
	\hline
	$E_7$ & $C_s - 810c_1^2S + 600c_1S^2 - 144S^3$ \\
	\hline
	$E_8$ & $C_s - 960c_1^2S + 840c_1S^2 - 240S^3$ \\
\end{tabular}
\end{center}
\caption{The values of $\pi_*(c_4(\tilde{X}_5))$ in the gravitational anomaly formula. $c_k$ denotes the $k^{\text{th}}$ Chern class of the base $B_4$. We set $C_s = 360c_1^3 + 12c_1c_2$ for convenience.}\label{t:Cs}
\end{table}

We test the gravitational anomaly cancellation using the series of CY fivefolds with non-abelian gauge groups we constructed in Section \ref{sec:CYseries}. The base manifolds are generalized Hirzebruch fourfolds $B_{n,4}$. The CY5s in this series all have non-Higgsable non-abelian gauge groups and have matter only in the adjoint representation of the gauge group. The anomaly can be cancelled when $c_1(B_4)\cdot\pi_*(G_4\cdot G_4)$ vanishes. For all the bases $B_{n,4}$, the divisor carrying non-Abelian gauge group is $S=\mb{P}^3$, and we have the purely geometric $\chi(\mathbf{adj})$:
\be
\ba
	\chi(\mathbf{adj}) &= -\frac{1}{24}\int_Sc_1(S)c_2(S) \\
	&= -1.
\ea
\ee

We summarize the topological numbers that are involved in the computation of gravitational anomaly cancellation in table \ref{t:Bn4-anomaly}. For completeness we also include in the table the case $B_{4,4}$ where there is no non-abelian gauge group. The bases in the table all have $\tau(B_{n,4}) = 0$ and $\chi_1(B_{n,4}) = -2$.
\begin{table}[h]
\begin{center}
\begin{tabular}{c|c|c|c|c}
	Base & $G$ & $\chi_1(\tilde{X}_5)$ & $\mathcal{A}_{\text{grav}|\text{3-7}}$ & $\mathcal{A}_{\text{grav}|\text{7-7}}$ \\
	\hline
	$B_{4,4}$ & - & 93\,188 & $-93\,216$ & 0 \\
	\hline
	$B_{6,4}$ & $SU(3)$ & 151\,466 & $-151\,488$ & $-6$ \\
	\hline
	$B_{8,4}$ & $SO(8)$ & 235\,292 & $-235\,296$ & $-24$ \\
	\hline
	$B_{12,4}$ & $E_6$ & 494\,924 & $-494\,880$ & $-72$ \\
	\hline
	$B_{24,4}$ & $E_8$ & 2\,314\,868 & $-2\,314\,656$ & $-240$
\end{tabular}
\end{center}
\caption{Gravitational anomaly coefficients for the generic elliptic fivefold over the generalized Hirzebruch fourfolds $B_{n,4}$.}\label{t:Bn4-anomaly}
\end{table}

Since the gravitational anomaly has already been cancelled, according to (\ref{eq:A37}) any consistent $G_4$-flux on these bases must satisfy the condition
\begin{align}
	c_1(B_4)\cdot\pi_*(G_4\cdot G_4) = 0.
\end{align}

For example, we consider $\tilde{X}_5\xrightarrow{\pi}B_{6,4}$ for which $h^{1,1}(\tilde{X}_5) = 5$. The five generators of $H^{1,1}(\tilde{X}_5)$ are two vertical divisors $D_1$ and $D_2$, two exceptional divisors $E_1$ and $E_2$ and the zero section $\sigma$ of the elliptic fibration. For simplicity we set:
\begin{equation}
	H_1 = D_1,\ H_2 = D_2,\ H_3 = E_1,\ H_4 = E_2,\ H_5 = \sigma.
\end{equation}
We consider the $G_4$-fluxes in the vertical cohomology group $H_V^{2,2}(X_5)$ therefore we have:
\begin{equation}
	G_4 = \sum_{i,j}n_{ij}H_i\cdot H_j.
\end{equation}
In order for $G_4$ to uplift to fluxes in F-theory, it must satisfy the transversality conditions:
\begin{align}
	\int G_4\wedge S_0\wedge \omega_4 = 0,\ \int G_4\wedge \omega_6 = 0,\ \forall\omega_4\in H^4(B_4),\ \omega_6\in H^6(B_4). \label{eq:trans}
\end{align}
We also require that $G_4$ does not break non-abelian gauge groups, therefore it satisfies:
\begin{equation}
	\int G_4\wedge E_i\wedge \omega_4 = 0,\ \forall\omega_4\in H^4(B_4)\,. \label{eq:nobreak}
\end{equation}
Applying (\ref{eq:trans}) and (\ref{eq:nobreak}) on $G_4$ using $\omega_4$'s generated by $H_i\cdot H_j$ and $\omega_6$'s generated by $H_i\cdot H_j\cdot H_k$, and making use of the intersection numbers on $B_{6,4}$, we have:
\be
\ba
	n_{11} &= 6 n_{13}+6 n_{31}-36 n_{33}, \\
	n_{22} &= -\frac{1}{6} n_{12}-\frac{1}{6} n_{13}-\frac{1}{6} n_{21}-\frac{1}{6}n_{31}+n_{33}, \\
   	n_{43} &= -\frac{1}{2} n_{13}+\frac{1}{3} n_{14}-\frac{1}{2} n_{31}+5 n_{33}-n_{34}+\frac{1}{3} n_{41}, \\
   	n_{44} &= \frac{1}{2} n_{13}-\frac{1}{6} n_{14}+\frac{1}{2} n_{31}-5 n_{33}-\frac{1}{6} n_{41}, \\
   	n_{52} &= -\frac{5}{6} n_{13}-n_{15}-n_{25}-\frac{5}{6} n_{31}+5 n_{33}-n_{51}, \\
   	n_{55} &= -\frac{1}{6} n_{13}-\frac{1}{6} n_{15}-\frac{1}{6}n_{31}+n_{33}-\frac{1}{6} n_{51}.
\ea
\ee
It is then easy to show that any $G_4 = \sum_{i,j}n_{ij}H_i\cdot H_j$ such that the $n_{ij}$'s satisfy the above restrictions also satisfies $c_1(B_{6,4})\cdot\pi_*(G_4\cdot G_4) = 0$.

Here we also prove that for an $X_5$ over a smooth $B_4$ with no gauge group (codimension-one singular fiber), we always have
\be
c_1(B_4)\cdot\pi_*(G_4\cdot G_4) = 0\,.
\ee
We prove the uplift of such equality in $X_5$:
\be
\pi^*(c_1(B_4))\cdot G_4\cdot G_4=0\,.\label{cGG-uplift}
\ee

Denote the zero section by $S_0$ and vertical divisors in $X_5$ by $D_i$, the general form of vertical $G_4$-flux is
\be
G_4=a S_0\cdot S_0+\sum_i b_i S_0\cdot D_i+\sum_{i,j} c_{ij}D_i\cdot D_j\,.\label{smooth-G4}
\ee
We write the anticanonical divisor of $B_4$ as 
\be
-K(B_4)=c_1(B_4)=\sum_i m_i \pi_*(D_i)\,,
\ee
where $m_i\in\mb{Z}$ are coefficients associated to $B_4$ and the choice of basis $D_i$.

Then we have the following intersection number relations in $X_5$:
\be
\ba
S_0^2\cdot D_i\cdot D_j\cdot D_k&=-S_0\cdot\pi^*(c_1(B_4))\cdot D_i\cdot D_j\cdot D_k\cr
&=-\sum_l m_l S_0\cdot D_i\cdot D_j\cdot D_k\cdot D_l\,.\label{X5-int-rel1}
\ea
\ee

\be
\ba
S_0^3\cdot D_i\cdot D_j&=-S_0^2\cdot\pi^*(c_1(B_4))\cdot D_i\cdot D_j\cr
&=\sum_{k,l} m_k m_l S_0\cdot D_i\cdot D_j\cdot D_k\cdot D_l\,.\label{X5-int-rel2}
\ea
\ee

\be
\ba
S_0^4\cdot D_i&=-S_0^3\cdot\pi^*(c_1(B_4))\cdot D_i\cr
&=-\sum_{j,k,l} m_j m_k m_l S_0\cdot D_i\cdot D_j\cdot D_k\cdot D_l\,.\label{X5-int-rel3}
\ea
\ee

Besides these relations, we have obviously $D_i\cdot D_j\cdot D_k\cdot D_l\cdot D_m=0$.

The transversality conditions (\ref{eq:trans}) on $G_4$ become:
\be
G_4\cdot S_0\cdot D_i\cdot D_j=0\ ,\ G_4\cdot D_i\cdot D_j\cdot D_k=0
\ee
for any $i$, $j$, $k$. Plug in (\ref{smooth-G4}) and using (\ref{X5-int-rel1}, \ref{X5-int-rel2}, \ref{X5-int-rel3}), they are further reduced to
\be
\sum_{k,l}(am_k m_l-b_k m_l+c_{k,l})S_0\cdot D_i\cdot D_j\cdot D_k\cdot D_l=0\,,
\ee
\be
\sum_l(am_l-b_l)S_0\cdot D_i\cdot D_j\cdot D_k\cdot D_l=0\,.
\ee

Note that these equations are equivalent to the following equations:
\be
\ba
\sum_l(am_l-b_l)S_0\cdot D_i\cdot D_j\cdot D_k\cdot D_l&=0\cr
\sum_{k,l}c_{k,l}S_0\cdot D_i\cdot D_j\cdot D_k\cdot D_l&=0\,.\label{trans-eqs}
\ea
\ee

Now we can rewrite (\ref{cGG-uplift}):
\be
\ba
\pi^*(c_1(B_4))\cdot G_4\cdot G_4&=\sum_l m_l D_l\cdot G_4\cdot G_4\cr
&=\sum_{i,j,k,l}m_l(-a^2 m_i m_j m_k+2a b_i m_j m_k-b_i b_j m_k-2 a c_{ij} m_k\cr
&+2b_i c_{jk})S_0\cdot D_i\cdot D_j\cdot D_k\cdot D_l\,.
\ea
\ee 
Due to the second equation of (\ref{trans-eqs}), the terms with $c_{ij}$ all vanishes. The rest of terms vanish as well from the first equation of (\ref{trans-eqs}).

Thus we have proved (\ref{cGG-uplift}) and equivalently
\be
c_1(B_4)\cdot\pi_*(G_4\cdot G_4) = 0\,.
\ee

For example, we can simply check that the generic fibration $X_5$ over $\mb{P}^4$, with Hodge numbers $(h^{1,1},h^{2,1},h^{3,1},h^{4,1})=(2,0,0,56\,977)$ satisfies the gravitional anomaly cancellation with $G_4$ flux. This also holds for $B_{4,4}$.

\subsection{Orbifold singularity and anomaly}

In this section, we consider a number of bases with orbifold singularity, and check the gravitational anomaly cancellation in these cases. As a result, we found that there need to be finite contributions from the orbifold singularities to cancel the anomaly.

The bases we considered are the weighted projective space $\mb{P}^{1,1,1,1,n}$, where $n$ takes the values in table~\ref{t:Bn}. The rays of the base are
\be
\ba
	&v_1 = (1,0,0,0),\ v_2 = (0,1,0,0),\ v_3 = (0,0,1,0),\ v_4 = (0,0,0,1), \\
	&v_5 = (-1,-1,-1,-n)\,.
\ea
\ee
The 4d cones are
\be
      (1,2,3,4),\ (1,2,3,5),\ (1,2,4,5),\ (1,3,4,5),\ (2,3,4,5)\,.
\ee

As the volume of the 4d cone Vol$(v_1 v_2 v_3 v_5)=n$, there is a $\mb{C}^4/\mb{Z}_n$ orbifold singularity at $z_1=z_2=z_3=z_5=0$. Unlike $B_n$, there is no toric divisor carrying non-Higgsable gauge group on $\mb{P}^{1,1,1,1,n}$. 

Now we compute the topological quantities involved in the gravitational anomaly cancellation (\ref{eq:A37}). For the divisors $D_i$ corresponds to the ray $v_i$, we have the linear equivalence relation:
\be
D_1=D_2=D_3=D_5=H\ ,\ D_4=nH\,,
\ee
and intersection number
\be
H^4=\frac{1}{n}\,.
\ee

The various Chern classes of $\mb{P}^{1,1,1,1,n}$ are
\be
\ba
c_1&=(n+4)H\cr
c_2&=(4n+6)H^2\cr
c_3&=(6n+4)H^3\cr
c_4&=4+\frac{1}{n}\,.
\ea
\ee

However, in this case the formula (\ref{tauB4}) and (\ref{chi1B4}) will no longer hold, as they give rise to fractional numbers for a general $n$. For a singular toric variety, the topological numbers $\tau$ and $\chi_1$ are computed in a combinatoric way instead \cite{maxim2015characteristic}. In particular, these numbers for $\mb{P}^{1,1,1,1,n}$ are exactly the same as the ones of $\mb{P}^4$:
\be
\ba
\tau(\mb{P}^{1,1,1,1,n})&=1\cr
\chi_1(\mb{P}^{1,1,1,1,n})&=-1\,.
\ea
\ee
Adding up the contributions in (\ref{eq:A37}), we get the total gravitational anomaly:
\be
\ba
\label{orbifold-grav-anomaly}
\mathcal{A}_{\rm grav}&=-\tau(\mb{P}^{1,1,1,1,n})+\chi_1(X_5)-2\chi_1(\mb{P}^{1,1,1,1,n})+24-\frac{1}{4}c_1\cdot (360 c_1^3+12 c_1 c_2)\cr
&=\chi_1(X_5)-90n^3-1\,452 n^2-8\,754 n-23\,351-\frac{23\,328}{n}\,.
\ea
\ee

On the other hand, the Hodge numbers of the smooth $X_5$ over $\mb{P}^{1,1,1,1,n}$ is the same as the generic elliptic CY5 over $B_n$, given in table~\ref{t:Bn}. The reason is that the 6d reflexive polytopes for $X_5$ are exactly the same in the two cases, and the Batyrev formula (\ref{Batyrevh11}, \ref{Batyrevhm1}, \ref{Batyrevhd1}) hold. Plug in the $\chi(X_5)$ from table~\ref{t:Bn} into (\ref{orbifold-grav-anomaly}), we found that $\mathcal{A}_{\rm grav}$ is always non-zero. To compensate this, we propose a new 2d sector from the orbifold singularity, which has the contribution $\mathcal{A}_{\rm orbifold}$ to $\mathcal{A}_{\rm grav}$ in table~\ref{t:orbifold-anomaly}. Notably, the case of $\mb{Z}_2$ and $\mb{Z}_3$ has a contribution $(-1)$ and $(+1)$, respectively. Hence a $\mb{C}^4/\mb{Z}_2$ singularity would effectively act as a Fermi or tensor multiplet, while a $\mb{C}^4/\mb{Z}_3$ singularity effectively acts as a chiral multiplet.

\begin{table}
\begin{center}
\begin{tabular}{|c|c|c|}
\hline
$\mb{Z}_n$ & $\mathcal{A}_{\rm orbifold}$\\
\hline
$\mb{Z}_2$ & $-1$\\
$\mb{Z}_3$ & $1$\\
$\mb{Z}_4$ & $3$\\
$\mb{Z}_6$ & $9$\\
$\mb{Z}_8$ & $15$\\
$\mb{Z}_{12}$ & $27$\\
$\mb{Z}_{24}$ & $63$\\
\hline
\end{tabular}
\caption{The additional gravitational anomaly contribution from $\mb{C}^4/\mb{Z}_n$ orbifold singularity on a compact singular base.}\label{t:orbifold-anomaly}
\end{center}
\end{table}

Finally we make more comments on the physics of singular bases in F-theory. In the case of singular base surface in 6d F-theory, such as the $\mb{Z}_3$ orbifold in~\cite{DelZotto:2014fia}, there is a localized SCFT sector coupled to gravity. The gravitational anomaly will cancel after the contribution of the SCFT is included. In the case of 2d F-theory here, we expect a similar story. Nonetheless, for the case of $\mb{Z}_2$ and $\mb{Z}_3$, one cannot blow up the singular loci and still get a $X_5$ with the same Hodge numbers. One can also check this from the Hodge numbers of $X_5$ in table~\ref{t:Bn}, where the $h^{1,1}(X_5)$ over $B_2$ and $B_3$ are the same as the ones over $\mb{P}^4$. The SCFT sector in these cases will not have a Coulomb branch after the dimensional reduction to 1d.

\section{Discussions}

In this paper, we constructed the elliptic fibration structure for a variety of elliptic Calabi-Yau fivefolds. Especially, we studied the elliptic Calabi-Yau fivefolds with the largest $h^{1,1}$ or $h^{4,1}$, as well as hypersurfaces of weighted projective spaces with small degrees. The non-vanishing $h^{2,1}$ and $h^{3,1}$ in some examples are explained as well. Nonetheless, we have not studied the detailed condition on $G_4$ flux in many of these geometries. For example, for the Calabi-Yau fivefold with the largest $h^{1,1}$, one needs to know whether a non-vanishing $G_4$ is required, and if a generic $G_4$ flux choice would break any gauge symmetry. This would be a question for the future work.

Moreover, one can also study the set of smooth compact fourfold bases in a more systematic way, applying the  methods in 4d F-theory, such as $\mb{P}^1$ fibrations~\cite{Halverson:2015jua}, Monte Carlo and random walk methods~\cite{Taylor:2015ppa,Taylor:2017yqr}, and systematic blow ups from weak-Fano bases~\cite{Halverson:2017ffz}. 

Besides the cases with a smooth fourfold base, we have also initiated the study of singular base fourfold in 2d F-theory. This question is especially interesting in 2d, because of the presence of pure gravitational anomaly and one can study the correction term from base singularities. In this paper, we have studied the contribution of an orbifold singularity $\mb{C}^4/\mb{Z}_n$ with $n=2,3,4,6,8,12,24$ in table~\ref{t:orbifold-anomaly}. For more general types of base singularities, we will study them in the future. Of course, it is also crucial to explain the physical origin of these effects. 

Finally, one can ask what are the details of the 2d (0,2) SCFT constructed from either a base singularity or a non-minimal loci. It would be curious to relate them with the existing 2d (0,2) literature~\cite{Melnikov:2012hk,Benini:2013cda,Gadde:2013sca,Benini:2013xpa,Jia:2014ffa,Bobev:2014jva,Gadde:2014ppa,Guo:2015gha,Benini:2015bwz,Franco:2016nwv,Franco:2015tna,Chen:2016tdd,Franco:2016qxh,Apruzzi:2016nfr,Franco:2016fxm,Franco:2017cjj,Couzens:2017nnr,Dedushenko:2017osi,Closset:2017yte,Franco:2018qsc,Dimofte:2019buf}.

\subsection*{Acknowledgements}

We thank Jim Halverson and Benjamin Sung for helpful discussions and Sakura Schafer-Nameki for useful suggestions at the final stage of this work. The work of JT is supported by a grant from the Simons Foundation (\#488569, Bobby Acharya). YW is supported by the ERC Consolidator Grant number 682608 ``Higgs bundles: Supersymmetric Gauge Theories and Geometry (HIGGSBNDL)".

\appendix

\section{Blow up of $4-E_8$ Collision}
\label{sec:4-E8-blp}

\begin{figure}
\begin{center}
\includegraphics[height=6cm]{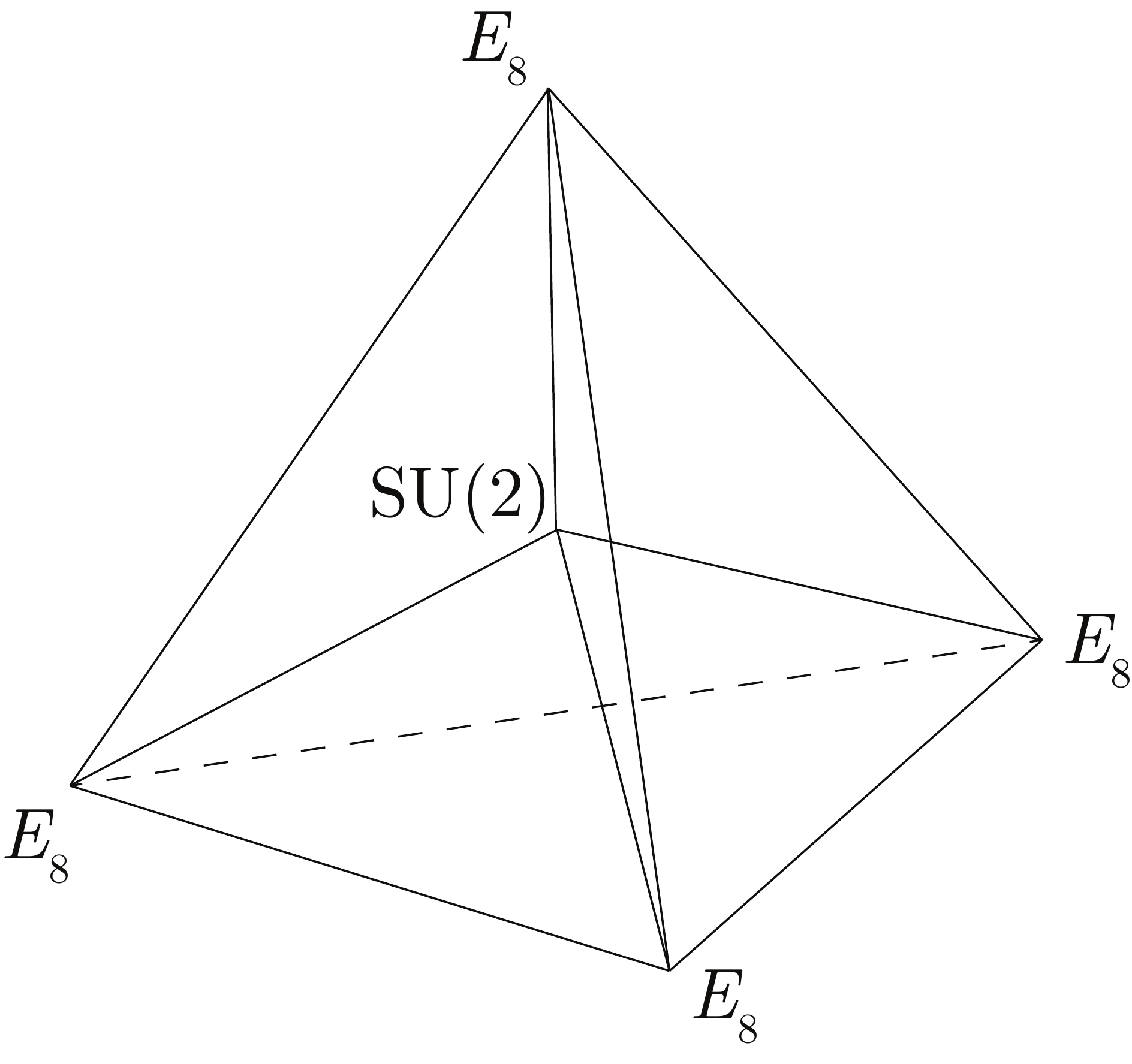}
\caption{The blow up of the intersection point of four divisors with $E_8$.}\label{f:4E8}
\end{center}
\end{figure}

Consider four rays $v_1,v_2,v_3,v_4$ of the fan corresponding to the toric base $B_4$, and for each of them we tune an $E_8$ gauge algebra along the toric divisor $D_i:= \{z_i = 0\}$ corresponding to it hence the order of vanishing of $(f,g)$ along $z_i$ is $(4,5)$ and each $D_i$ carries Kodaira type $II^*$ fiber. To remove the non-minimal loci, we first blow up the intersection point $z_1=z_2=z_3=z_4=0$ by adding a new ray $v_E = v_1+v_2+v_3+v_4$ that corresponds to the exceptional divisor $D_E$ of the blow-up. Using the linearity of the inner product it is easy to show that the order of vanishing of $(f,g)$ along $D_E$ is $(4,2)$. Therefore $D_E$ carries Kodaira type $IV$ fiber and supports an $SU(2)$ gauge group. The configuration of the base structure is plotted in figure~\ref{f:4E8} after the blow-up.

Then there are six new 3d cones with $(E_8,E_8,SU(2))$ Collisions: 
\be
\ba
&(v_1,v_2,v_1+v_2+v_3+v_4),(v_1,v_3,v_1+v_2+v_3+v_4),(v_1,v_4,v_1+v_2+v_3+v_4),\cr
&(v_2,v_3,v_1+v_2+v_3+v_4),(v_2,v_4,v_1+v_2+v_3+v_4),(v_3,v_4,v_1+v_2+v_3+v_4)\,.
\ea
\ee

\begin{figure}[h]
\begin{center}
\includegraphics[height=4cm]{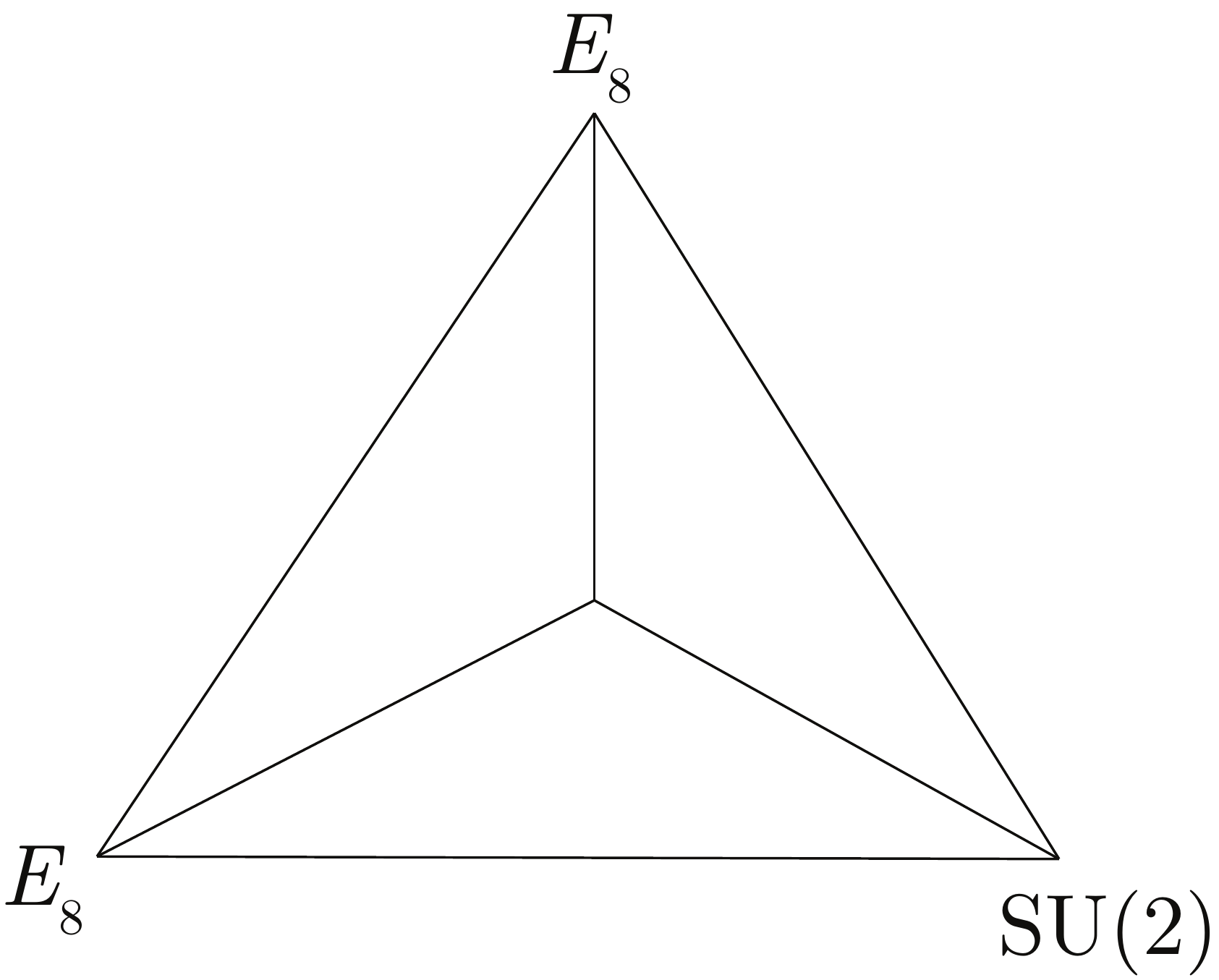}
\caption{The $(E_8,E_8,SU(2))$ collision and a single blow up at the intersection point.}\label{f:E8E8SU2}
\end{center}
\end{figure}

\begin{figure}[h]
\begin{center}
\includegraphics[height=6cm]{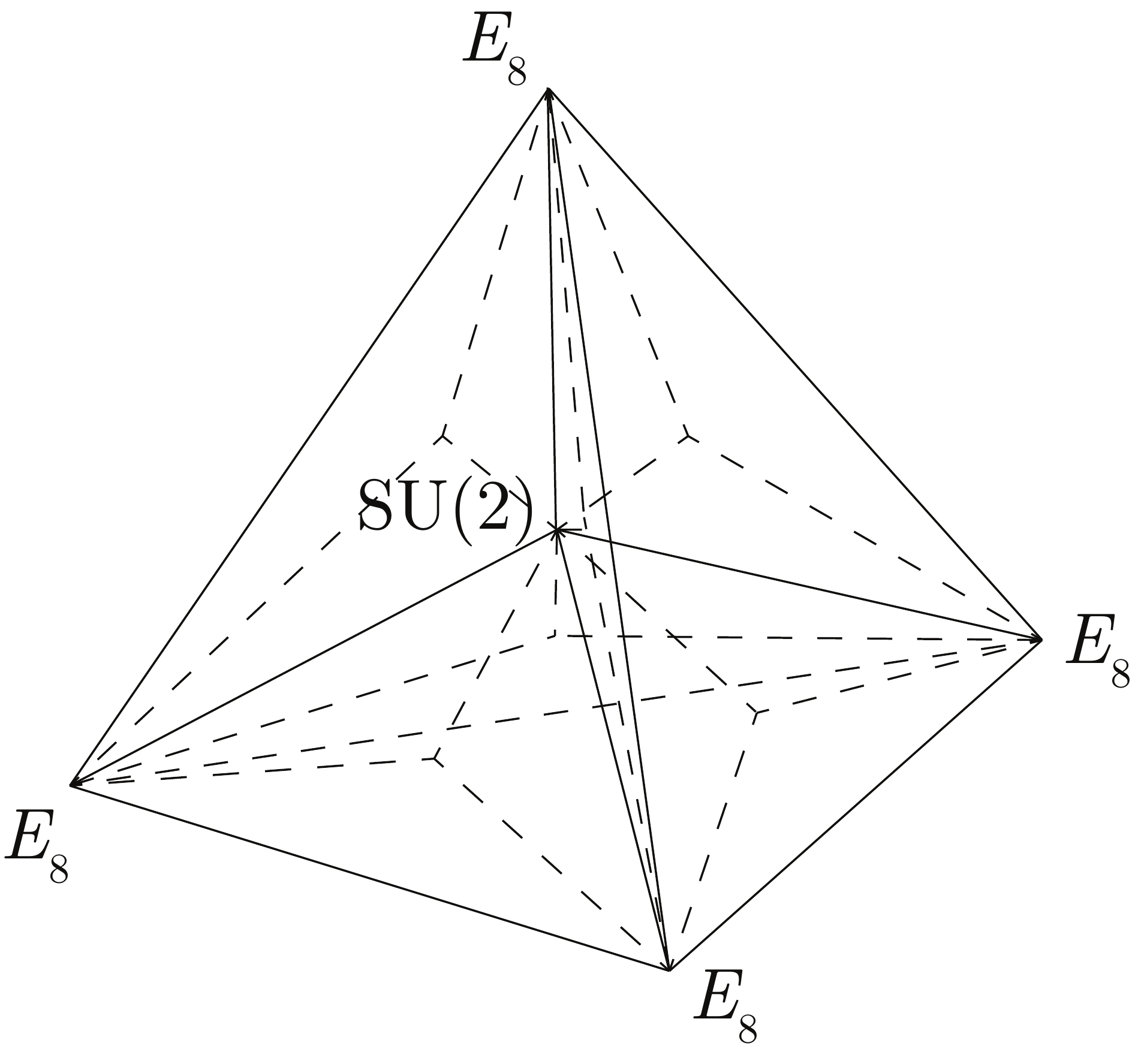}
\caption{The tetrahedron after the blow up of the $(E_8,E_8,E_8,E_8)$ and $(E_8,E_8,SU(2))$ collisions.}\label{f:4E8-2}
\end{center}
\end{figure}

Then we can blow up these codimension-three loci, according to figure~\ref{f:E8E8SU2}. Note that the order of vanishing of $g$ on the new exceptional divisor is zero. The whole tetrahedron will look like figure~\ref{f:4E8-2} after this step (we did not draw out all the subdivision cones).

\begin{figure}[h]
\begin{center}
\includegraphics[height=13cm]{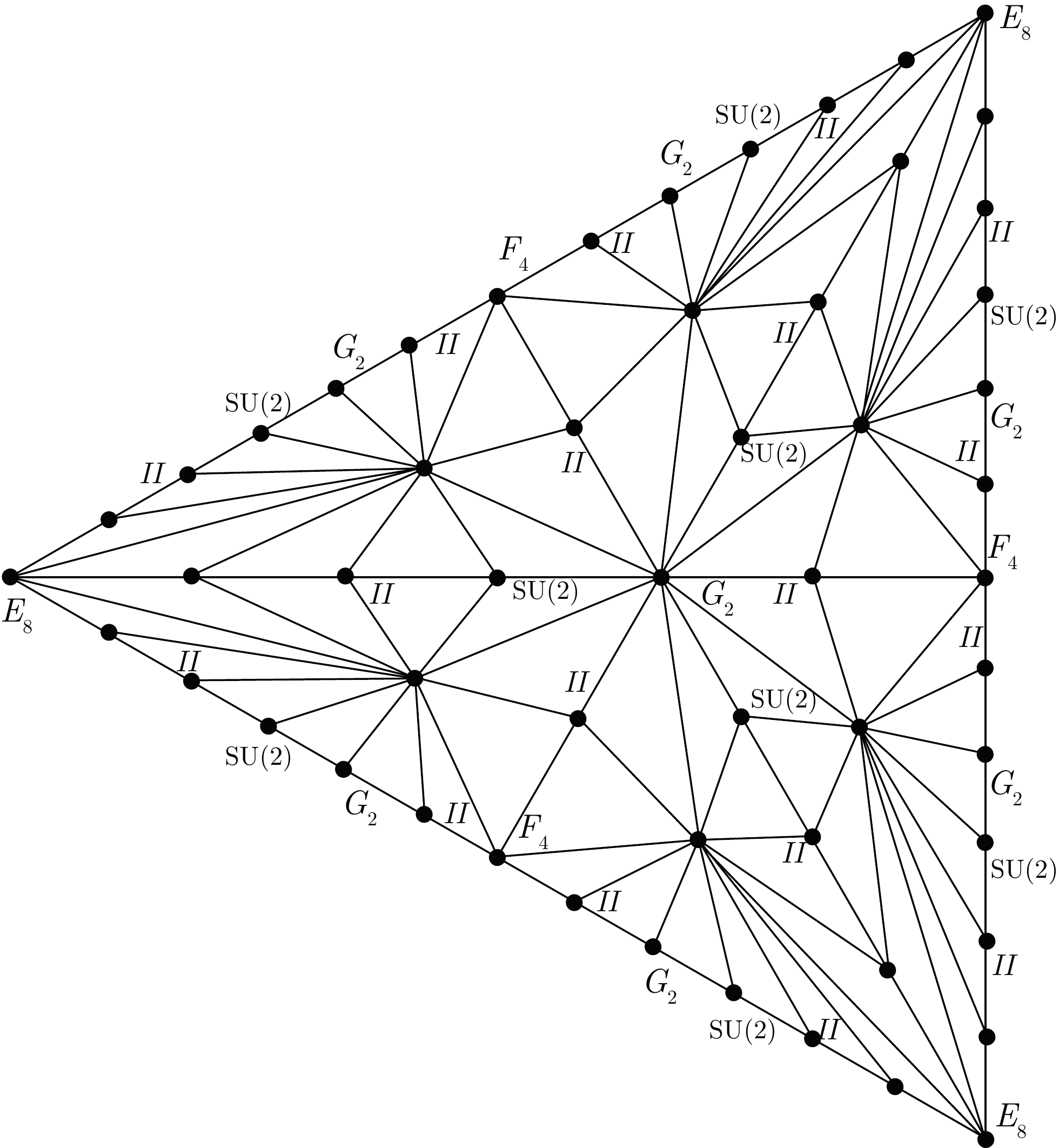}
\caption{The fully blown up $(E_8,E_8,E_8)$ collision.}\label{f:E8E8E8}
\end{center}
\end{figure}

\begin{figure}[h]
\begin{center}
\includegraphics[height=10cm]{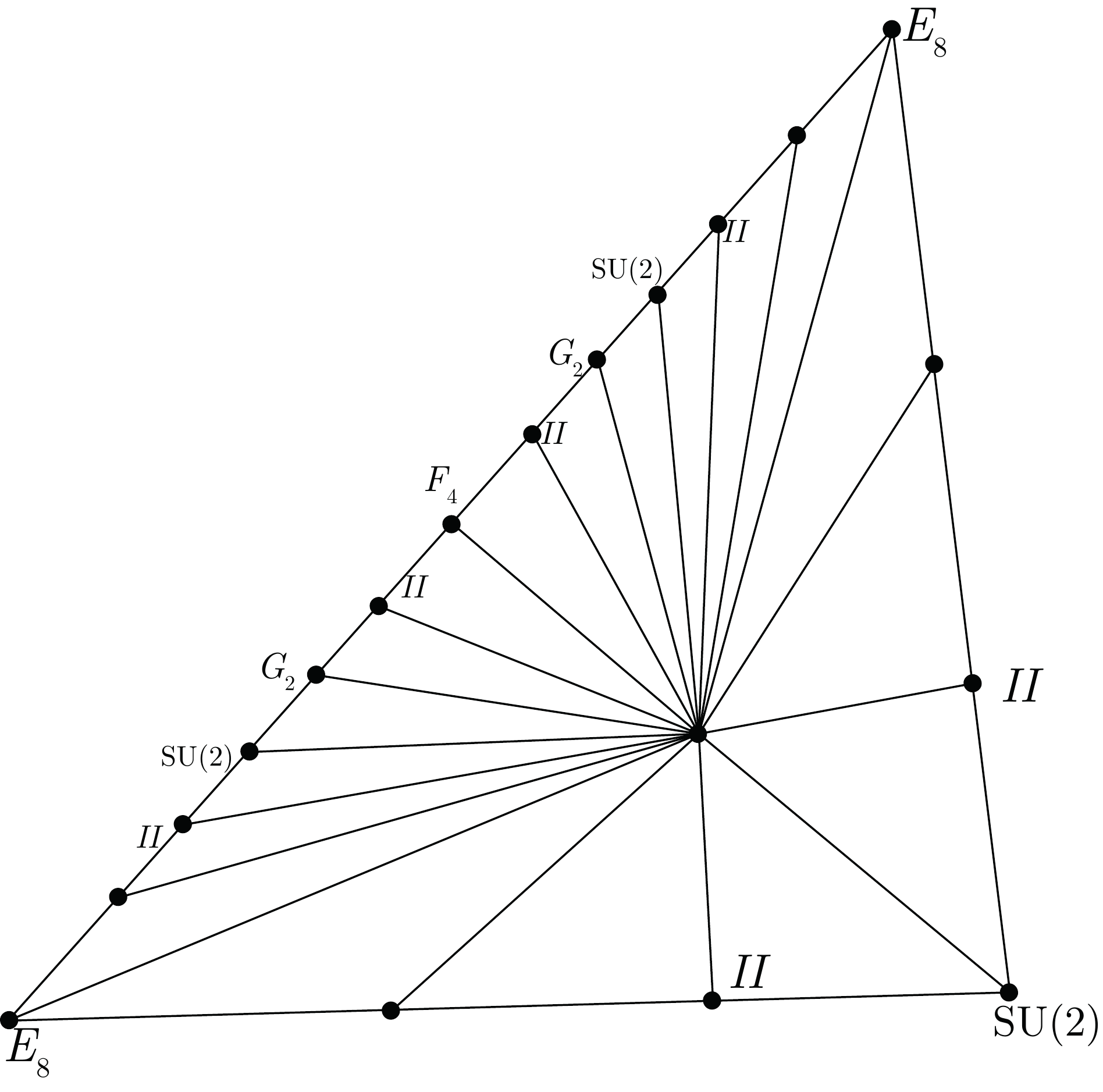}
\caption{The fully blown up $(E_8,E_8,SU(2))$ collision.}\label{f:E8E8SU2-2}
\end{center}
\end{figure}

After these blow ups, there are four $(E_8,SU(2))$ collisions in the middle of the tetrahedron, which need to be blown up twice for each. Finally, we can just blow up the four $(E_8,E_8,E_8)$ collisions on the faces, according to ~\cite{Wang:2020gmi}. For convenience purpose, we plot the blow up of $(E_8,E_8,E_8)$ collisions in figure~\ref{f:E8E8E8}.  We also show the fully subdivided $(E_8,E_8,SU(2))$ collision in figure~\ref{f:E8E8SU2-2}.
The final geometric configuration will be absent of non-minimal loci, and the elliptic fibration is flat.

\section{A list of elliptic Calabi-Yau fivefolds}
\label{app:CY5-list}

In this appendix, we list the elliptic CY5 as hypersurfaces of weighted projective spaces $\mb{P}^{1,w_1,w_2,w_3,w_4,w_5,w_6}$. We impose the following conditions:
\begin{enumerate}
\item{The lattice polytope associated to $\mb{P}^{1,w_1,w_2,w_3,w_4,w_5,w_6}$ is reflexive.}
\item{The weights satisfy $3w_5=2w_6$ and $1+w_1+w_2+w_3+w_4+w_5=w_6$, such that the 6d rays can be rotated by the matrix \ref{SL6rotation} to get a $\mb{P}^{2,3,1}$ fibration structure (the ``naive'' piling).}
\end{enumerate}

To get a finite list, we require that the degree\footnote{The complete list of weights giving rise to reflexive polytopes in this category was already worked out in~\cite{Kreuzer:2001fu}.}
\be
d\equiv 1+w_1+w_2+w_3+w_4+w_5+w_6\leq 150\,.
\ee

We list these models along with the Hodge numbers of CY5 and the 2d F-theory geometric gauge group in table~\ref{t:CY5-1}--\ref{t:CY5-4}. Note that for many cases, the base fourfold has to be singular. We do not list all the possible base topologies in detail.

\begin{table}
\begin{center}
\begin{tabular}{|c|c|c|c|}
\hline
$(w_1,w_2,w_3,w_4,w_5,w_6)$ & $(h^{1,1},h^{2,1},h^{3,1},h^{4,1},h^{2,3})$ &  Gauge group \\
\hline
$(1,1,1,1,10,15)$ & $(2,0,0,56\,977, 626\,727)$ &  -- \\
\hline
$(1, 1, 1, 2, 12, 18)$ & $(2, 0, 0, 59\,054, 649\,574)$ & -- \\
\hline
$(1, 1, 1, 3, 14, 21)$ & $(2, 0, 0, 72\,888, 801\,751)$ &  -- \\
\hline
$(1, 1, 2, 2, 14, 21)$ & $(2, 0, 0, 54\,703, 601\,696)$ & -- \\
\hline
$(1, 1, 1, 4, 16, 24)$ & $(3, 1, 0, 93\,190, 1\,025\,070)$  & -- \\
\hline
$(1, 1, 2, 3, 16, 24)$ & $(2, 0, 0, 62\,187, 684\,040)$ & -- \\
\hline
$(1, 2, 2, 2, 16, 24)$ & $(3, 0, 0, 46\,727, 513\,968)$ &  -- \\
\hline
$(1, 1, 3, 3, 18, 27)$ & $(3, 0, 2, 66\,398, 730\,358)$ & -- \\
\hline
$(1, 2, 2, 3, 18, 27)$ & $(2, 0, 0, 49\,821, 547\,988)$ & -- \\
\hline
$(2, 2, 2, 2, 18, 27)$ & $(4, 0, 0, 39\,495,434\,405)$ & $G_2$ \\
\hline
$(1, 1, 1, 6, 20, 30)$ & $(5, 0, 0, 151\,471, 1\,666\,132)$ & $SU(3)$ \\
\hline
$(1, 1, 2, 5, 20, 30)$ & $(3, 1, 0, 90\,996, 1\,000\,936)$ & -- \\
\hline
$(1, 1, 3, 4, 20, 30)$ & $(2, 0, 0, 75\,877, 834\,635)$ & -- \\
\hline
$(1, 2, 2, 4, 20, 30)$ & $(3, 0, 0, 56\,975, 626\,700)$ & -- \\
\hline
$(1, 2, 3, 3, 20, 30)$ & $(2, 0, 0, 50\,618, 556\,783)$ & -- \\
\hline
$(2, 2, 2, 3, 20, 30)$ & $(3, 0, 0, 38\,076, 418\,810)$ & -- \\
\hline
$(1, 1, 2, 6, 22, 33)$ & $(3, 0, 0, 110\,963, 1\,220\,566)$ & -- \\
\hline
$(1, 3, 3, 3, 22, 33)$ & $(4, 0, 45, 49\,405, 543\,298)$ & $SU(2)$ \\
\hline
$(2, 2, 3, 3, 22, 33)$ & $(2, 0, 0, 37\,069, 407\,745)$ & -- \\
\hline
$(1, 1, 1, 8, 24, 36)$ & $(7, 0, 0, 235\,299, 2\,588\,220)$ & $SO(8)$ \\
\hline
$(1, 1, 3, 6, 24, 36)$ & $(4, 1, 1, 104\,806, 1\,152\,847)$ & -- \\
\hline
$(1, 2, 2, 6, 24, 36)$ & $(4, 1, 0, 78\,669, 865\,332)$ & -- \\
\hline
$(1, 2, 4, 4, 24, 36)$ & $(4, 0, 2, 59\,052, 649\,543)$ & -- \\
\hline
$(1, 3, 3, 4, 24, 36)$ & $(3, 0, 3, 52\,470, 577\,153)$ & -- \\
\hline
$(2, 2, 3, 4, 24, 36)$ & $(3, 0, 0, 39\,435, 433\,760)$ & -- \\
\hline
$(2, 3, 3, 3, 24, 36)$ & $(4, 0, 0, 35\,110, 386\,173)$ & -- \\
\hline
$(1, 2, 3, 6, 26, 39)$ & $(2, 0, 0, 72\,216, 794\,363)$ & -- \\
\hline
$(2, 2, 2, 6, 26, 39)$ & $(4, 0, 0, 56\,282, 619\,065)$ & $G_2$ \\
\hline
$(3, 3, 3, 3, 26, 39)$ & $(6, 0, 0, 33\,809, 371\,839)$ & $F_4$ \\
\hline
$(1, 1, 4, 7, 28, 42)$ & $(4, 2, 0, 124\,796, 1\,372\,743)$ & -- \\
\hline
$(1, 2, 3, 7, 28, 42)$ & $(3, 1, 0, 83\,223, 915\,441)$ & -- \\
\hline
$(1, 2, 4, 6, 28, 42)$ & $(3, 0, 0, 72\,887, 801\,738)$ & -- \\
\hline
$(1, 3, 3, 6, 28, 42)$ & $(4, 0, 36, 64\,770, 712\,343)$ & $SU(2)$ \\
\hline
$(1, 4, 4, 4, 28, 42)$ & $(5, 0, 93, 54\,700, 601\,299)$ & $G_2$ \\
\hline
$(2, 2, 2, 7, 28, 42)$ & $(5, 2, 0, 62\,633, 688\,936)$ & -- \\
\hline
$(2, 2, 3, 6, 28, 42)$ & $(3, 0, 0, 48\,652, 535\,150)$ & -- \\
\hline
$(2, 3, 4, 4, 28, 42)$ & $(3, 0, 0, 36\,516, 401\,654)$ & -- \\
\hline
$(3, 3, 3, 4, 28, 42)$ & $(4, 0, 78, 32\,417, 356\,321)$ & -- \\
\hline
$(1, 1, 2, 10, 30, 45)$ & $(7, 0, 0, 229\,774, 2\,527\,448)$ & $SO(8)$ \\
\hline
$(1, 1, 3, 9, 30, 45)$ & $(6, 0, 1, 170\,371, 1\,874\,032)$ & $SU(2)$\\
\hline
$(1, 1, 6, 6, 30, 45)$ & $(5, 13, 0, 127\,933, 1\,407\,347)$ & $G_2$ \\
\hline
\end{tabular}
\caption{A list of elliptic CY5 hypersurfaces in weighted projective spaces $\mb{P}^{1,w_1,w_2,w_3,w_4,w_5,w_6}$.}\label{t:CY5-1}
\end{center}
\end{table}

\begin{table}
\begin{center}
\begin{tabular}{|c|c|c|c|}
\hline
$(w_1,w_2,w_3,w_4,w_5,w_6)$ & $(h^{1,1},h^{2,1},h^{3,1},h^{4,1},h^{2,3})$ &  Gauge group \\
\hline
$(1, 2, 2, 9, 30, 45)$ & $(5, 0, 0, 127\,801, 1\,405\,768)$ & $SU(3)$\\
\hline
$(1, 2, 5, 6, 30, 45)$ & $(3, 1, 0, 76\,807, 844\,867)$ & -- \\
\hline
$(1, 3, 5, 5, 30, 45)$ & $(4, 0, 4, 61\,469, 676\,141)$ & -- \\
\hline
$(2, 2, 5, 5, 30, 45)$ & $(4, 0, 4, 46\,113, 507\,225)$ & -- \\
\hline
$(2, 3, 3, 6, 30, 45)$ & $(4, 0, 0, 42\,783, 470\,581)$ & -- \\
\hline
$(3, 3, 3, 5, 30, 45)$ & $(5, 1, 0, 34\,351, 377\,825)$ & --\\
\hline
$(1, 1, 1, 12, 32, 48)$ & $(9, 0, 0, 494\,933, 5\,444\,174)$ & $E_6$ \\
\hline
$(1, 2, 4, 8, 32, 48)$ & $(5, 1, 1, 93\,189, 1\,025\,052)$ & --\\
\hline
$(1, 2, 6, 6, 32, 48)$ & $(5, 7, 0, 82\,874, 911\,639)$ & $SU(2)$ \\
\hline
$(1, 3, 3, 8, 32, 48)$ & $(4, 2, 0, 82\,820, 911\,007)$ & -- \\
\hline
$(1, 4, 4, 6, 32, 48)$ & $(4, 0, 3, 62\,185, 684\,012)$ & -- \\
\hline
$(2, 2, 3, 8, 32, 48)$ & $(4, 1, 0, 62\,184, 684\,002)$ & -- \\
\hline
$(2, 3, 4, 6, 32, 48)$ & $(3, 0, 0, 41\,498, 456\,458)$ & -- \\
\hline
$(3, 3, 3, 6, 32, 48)$ & $(6, 0, 0, 38\,407, 422\,421)$ & $F_4$ \\
\hline
$(3, 4, 4, 4, 32, 48)$ & $(5, 0, 0, 31\,266, 343\,882)$ & $SU(2)$ \\
\hline
$(1, 3, 6, 6, 34, 51)$ & $(5, 0, 28, 70\,409, 774\,390)$ & -- \\
\hline
$(2, 2, 6, 6, 34, 51)$ & $(5, 0, 0, 54\,381, 598\,162)$ & $G_2$ \\
\hline
$(1, 1, 3, 12, 36, 54)$ & $(8, 0, 1, 264\,675, 2\,911\,356)$ & $SO(8)$ \\
\hline
$(1, 1, 6, 9, 36, 54)$ & $(6, 3, 0, 176\,763, 1\,944\,376)$ & --\\
\hline
$(1, 2, 2, 12, 36, 54)$ & $(8, 0, 0, 198\,578, 2\,184\,285)$ & $SO(8)$ \\
\hline
$(1, 3, 4, 9, 36, 54)$ & $(4, 1, 1, 88\,430, 972\,719)$ & -- \\
\hline
$(1, 4, 6, 6, 36, 54)$ & $(5, 0, 4, 66\,397, 730\,342)$ & -- \\
\hline
$(2, 2, 4, 9, 36, 54)$ & $(5, 2, 0, 66\,499, 731\,468)$ & --  \\
\hline
$(2, 3, 3, 9, 36, 54)$ & $(5, 1, 0, 59\,050, 649\,517)$ & -- \\
\hline
$(2, 3, 6, 6, 36, 54)$ & $(6, 0, 2, 44\,360, 487\,913)$ & -- \\
\hline
$(2, 5, 5, 5, 36, 54)$ & $(7, 0, 0, 38\,221, 421\,070)$ & $G_2$  \\
\hline
$(3, 4, 4, 6, 36, 54)$ & $(4, 0, 2, 33\,242, 365\,639)$ & -- \\
\hline
$(3, 3, 6, 6, 38, 57)$ & $(6, 0, 0, 37\,938, 417\,281)$ & $F_4$ \\
\hline
$(1, 1, 2, 15, 40, 60)$ & $(9, 0, 0, 483\,320, 5\,316\,436)$ & $E_6$ \\
\hline
$(1, 1, 5, 12, 40, 60)$ & $(7, 1, 0, 242\,246, 2\,664\,655)$ & $SU(3)$ \\
\hline
$(1, 2, 4, 12, 40, 60)$ & $(7, 0, 1, 151\,470, 1\,666\,115)$ & $SU(3)$ \\
\hline
$(1, 2, 6, 10, 40, 60)$ & $(5, 2, 0, 121\,291, 1\,334\,182)$ & --  \\
\hline
$(1, 2, 8, 8, 40, 60)$ & $(6, 13, 0, 113\,741, 1\,251\,228)$ & $G_2$ \\
\hline
$(1, 3, 3, 12, 40, 60)$ & $(7, 0, 36, 134\,621, 1\,480\,680)$ & $SU(3)$ \\
\hline
$(1, 3, 5, 10, 40, 60)$ & $(5, 1, 2, 97\,033, 1\,067\,349)$ & -- \\
\hline
$(1, 4, 4, 10, 40, 60)$ & $(6, 2, 4, 90\,994, 1\,000\,909)$ & -- \\
\hline
$(1, 4, 6, 8, 40, 60)$ & $(4, 0, 2, 75\,876, 834\,617)$ & -- \\
\hline
$(1, 5, 5, 8, 40, 60)$ & $(5, 0, 6, 72\,831, 801\,120)$ & -- \\
\hline
$(1, 6, 6, 6, 40, 60)$ & $(7, 0, 72, 67\,467, 741\,894)$ & $F_4$ \\

\hline
\end{tabular}
\caption{A list of elliptic CY5 hypersurfaces in weighted projective spaces $\mb{P}^{1,w_1,w_2,w_3,w_4,w_5,w_6}$ (cont.).}\label{t:CY5-2}
\end{center}
\end{table}

\begin{table}
\begin{center}
\begin{tabular}{|c|c|c|c|}
\hline
$(w_1,w_2,w_3,w_4,w_5,w_6)$ & $(h^{1,1},h^{2,1},h^{3,1},h^{4,1},h^{2,3})$ &  Gauge group \\
\hline
$(2, 2, 3, 12, 40, 60)$ & $(6, 0, 0, 101\,041, 1\,111\,401)$ & $SU(3)$ \\
\hline
$(2, 2, 5, 10, 40, 60)$ & $(6, 1, 2, 72\,861, 801\,444)$ & -- \\
\hline
$(2, 3, 4, 10, 40, 60)$ & $(4, 1, 0, 60\,700, 667\,682)$ & -- \\
\hline
$(2, 3, 6, 8, 40, 60)$ & $(3, 0, 0, 50\,616, 556\,764)$ & -- \\
\hline
$(2, 4, 5, 8, 40, 60)$ & $(4, 0, 2, 45\,580, 501\,359)$ & -- \\
\hline
$(2, 5, 6, 6, 40, 60)$ & $(4, 1, 0, 40\,533, 445\,844)$ & -- \\
\hline
$(3, 3, 3, 10, 40, 60)$ & $(7, 3, 171, 53\,932, 592\,648)$ & $SU(2)$ \\
\hline
$(3, 3, 5, 8, 40, 60)$ & $(3, 1, 0, 40\,478, 445\,255)$ & -- \\
\hline
$(3, 4, 4, 8, 40, 60)$ & $(5, 0, 0, 38\,072, 418\,756)$ & -- \\
\hline
$(3, 4, 6, 6, 40, 60)$ & $(5, 0, 36, 33\,766, 371\,296)$ & $SU(2)$ \\
\hline
$(3, 5, 5, 6, 40, 60)$ & $(4, 0, 6, 32\,395, 356\,333)$ & -- \\
\hline
$(4, 4, 5, 6, 40, 60)$ & $(4, 0, 4, 30\,404, 334\,425)$ & --\\
\hline
$(4, 5, 5, 5, 40, 60)$ & $(7, 1, 0, 29\,366, 322\,971)$ & -- \\
\hline
$(1, 1, 9, 9, 42, 63)$ & $(7, 12, 0, 218\,302, 2\,401\,378)$ & $F_4$\\
\hline
$(1, 2, 3, 14, 42, 63)$ & $(7, 0, 0, 210\,158, 2\,311\,680)$ & $SO(8)$\\
\hline
$(1, 3, 7, 9, 42, 63)$ & $(4, 1, 2, 93\,637, 1\,029\,994)$ & -- \\
\hline
$(1, 6, 6, 7, 42, 63)$ & $(5, 19, 0, 70\,261, 773\,021)$ & $G_2$ \\
\hline
$(2, 2, 2, 14, 42, 63)$ & $(10, 0, 0, 161\,415, 1\,775\,471)$ & $G_2$, $SO(8)$\\
\hline
$(2, 2, 7, 9, 42, 63)$ & $(4, 2, 0, 70\,240, 772\,635)$ & -- \\
\hline
$(2, 3, 6, 9, 42, 63)$ & $(4, 0, 0, 54\,700, 601\,677)$ & -- \\
\hline
$(2, 6, 6, 6, 42, 63)$ & $(8, 0, 93, 42\,004, 461\,614)$ & $G_2$, $G_2$\\
\hline
$(3, 3, 7, 7, 42, 63)$ & $(6, 0, 12, 40\,161, 441\,751)$ & -- \\
\hline
$(1, 2, 6, 12, 44, 66)$ & $(6, 5, 0, 147\,911, 1\,627\,018)$ & $SU(2)$  \\
\hline
$(1, 3, 6, 11, 44, 66)$ & $(5, 3, 0, 107\,617, 1\,183\,779)$ & --  \\
\hline
$(1, 4, 4, 12, 44, 66)$ & $(6, 0, 75, 110\,961, 1\,220\,238)$ & $G_2$ \\
\hline
$(2, 2, 6, 11, 44, 66)$ & $(6, 3, 0, 80\,885, 889\,716)$ & -- \\
\hline
$(2, 3, 4, 12, 44, 66)$ & $(4, 0, 0, 74\,014, 814\,125)$ & -- \\
\hline
$(2, 4, 4, 11, 44, 66)$ & $(6, 3, 0, 60\,687, 667\,536)$ & -- \\
\hline
$(3, 3, 3, 12, 44, 66)$ & $(7, 0, 0, 67\,767, 745\,372)$ & $F_4$\\
\hline
$(3, 3, 4, 11, 44, 66)$ & $(4, 2, 0, 53\,832, 592\,147)$ & -- \\
\hline
$(3, 6, 6, 6, 44, 66)$ & $(9, 0, 45, 33\,939, 373\,123)$ & $F_4$, $SU(2)$ \\
\hline
$(1, 1, 3, 18, 48, 72)$ & $(10, 0, 1, 556\,757, 6\,124\,238)$ & $E_6$ \\
\hline
$(1, 1, 9, 12, 48, 72)$ & $(9, 3, 0, 279\,225, 3\,071\,429)$ & $SU(3)$ \\
\hline
$(1, 2, 2, 18, 48, 72)$ & $(10, 0, 0, 417\,652, 4\,594\,081)$ & $E_6$ \\
\hline
$(1, 2, 4, 16, 48, 72)$ & $(9, 0, 1, 235\,298, 2\,588\,205)$ & $SO(8)$ \\
\hline
$(1, 2, 8, 12, 48, 72)$ & $(7, 3, 0, 157\,144, 1\,728\,561)$ & --\\
\hline
$(1, 3, 3, 16, 48, 72)$ & $(9, 0, 7, 209\,132, 2\,300\,383)$ & $SO(8)$\\
\hline
$(1, 4, 6, 12, 48, 72)$ & $(7, 1, 3, 104\,805, 1\,152\,828)$ & -- \\
\hline
$(1, 4, 9, 9, 48, 72)$ & $(6, 8, 0, 93\,192, 1\,025\,143)$ & $SU(2)$ \\
\hline
$(1, 6, 8, 8, 48, 72)$ & $(7, 1, 6, 78\,672, 865\,365)$ & --\\
\hline
\end{tabular}
\caption{A list of elliptic CY5 hypersurfaces in weighted projective spaces $\mb{P}^{1,w_1,w_2,w_3,w_4,w_5,w_6}$ (cont.).}
\label{t:CY5-3}
\end{center}
\end{table}

\begin{table}
\begin{center}
\begin{tabular}{|c|c|c|c|}
\hline
$(w_1,w_2,w_3,w_4,w_5,w_6)$ & $(h^{1,1},h^{2,1},h^{3,1},h^{4,1},h^{2,3})$ &  Gauge group \\
\hline
$(2, 2, 3, 16, 48, 72)$ & $(8, 0, 0, 156\,930, 1\,726\,165)$ & $SO(8)$  \\
\hline
$(2, 3, 6, 12, 48, 72)$ & $(7, 1, 1, 69\,955, 769\,462)$ & $SU(2)$ \\
\hline
$(2, 3, 9, 9, 48, 72)$ & $(6, 7, 0, 62\,185, 684\,056)$ & $SU(2)$ \\
\hline
$(2, 4, 8, 9, 48, 72)$ & $(5, 1, 2, 52\,513, 577\,627)$ & --  \\
\hline
$(2, 6, 6, 9, 48, 72)$ & $(6, 0, 3, 46\,724, 513\,926)$ & --  \\
\hline
$(3, 3, 8, 9, 48, 72)$ & $(6, 2, 0, 46\,771, 514\,455)$ & -- \\
\hline
$(3, 4, 4, 12, 48, 72)$ & $(7, 1, 1, 52\,531, 577\,798)$ & -- \\
\hline
$(3, 4, 8, 8, 48, 72)$ & $(7, 0, 4, 39\,431, 433\,696)$ & -- \\
\hline
$(3, 6, 6, 8, 48, 72)$ & $(7, 1, 3, 35\,109, 386\,156)$ & -- \\
\hline
$(4, 4, 6, 9, 48, 72)$ & $(5, 0, 6, 35\,022, 385\,218)$ & -- \\
\hline
$(5, 5, 5, 8, 48, 72)$ & $(9, 2, 0, 30\,163, 333\,037)$ & $G_2$, $SU(2)$ \\
\hline
$(1, 2, 6, 15, 50, 75)$ & $(6, 0, 0, 197\,152, 2\,168\,627)$ & $SU(3)$ \\
\hline
$(1, 3, 5, 15, 50, 75)$ & $(7, 0, 2, 157\,742, 1\,735\,118)$ & $SU(3)$ \\
\hline
$(1, 3, 10, 10, 50, 75)$ & $(6, 13, 0, 118\,451, 1\,303\,044)$ & $G_2$\\
\hline
$(2, 2, 5, 15, 50, 75)$ & $(7, 0, 2, 118\,321, 1\,301\,488)$ & $SU(3)$ \\
\hline
$(2, 2, 10, 10, 50, 75)$ & $(8, 13, 0, 90\,632, 997\,010)$ & $G_2$, $G_2$ \\
\hline
$(2, 6, 6, 10, 50, 75)$ & $(5, 0, 0, 50\,391, 554\,292)$ & $G_2$ \\
\hline
$(3, 3, 3, 15, 50, 75)$ & $(10, 0, 0, 90\,009, 990\,007)$ & $F_4$, $SU(3)$ \\
\hline
$(3, 5, 6, 10, 50, 75)$ & $(4, 0, 4, 39\,526, 434\,773)$ & -- \\
\hline
$(6, 6, 6, 6, 50, 75)$ & $(11, 0, 2\,024, 28\,575, 303\,200)$ & $E_8$\\
\hline
\end{tabular}
\caption{A list of elliptic CY5 hypersurfaces in weighted projective spaces $\mb{P}^{1,w_1,w_2,w_3,w_4,w_5,w_6}$ (cont.).}
\label{t:CY5-4}
\end{center}
\end{table}

\newpage
\bibliographystyle{JHEP}
\bibliography{F-ref}

\end{document}